\documentclass[preprint2]{aastex}
\usepackage{epsf}
\shorttitle{Spectroscopy of M81 Globular Clusters}
\shortauthors{Schroder et al.}
\def\etal{et al.}
\def\bonafide{{\it bona fide}}

\def\kpc{{\rm\,kpc}}

\def\msun{{\rm\,M_\odot}}
\def\lsun{{\rm\,L_\odot}}

\begin{document}

\newcommand{\expon}[1]{\times \, 10^{#1}}
\newcommand{\myfig}[1]{Figure~\ref{#1}}
\newcommand{\mytab}[1]{Table~\ref{#1}}
\newcommand{\vrot}[1]{$\rm V_{rot}\!\!=\!\! #1 km \, s^{-1}$}
\newcommand{\BC}{Bruzual and Charlot}
\newcommand{\speedunit}{ $\rm \! km \, s^{-1}$}
\newcommand{\degree}{$^{\rm o}$}
\newcommand{\speed}[1]{$#1 \rm \, km \, s^{-1}$}
\newcommand{\speedpm}[2]{$#1\pm#2{\rm\,km \, s^{-1}}$}
\newcommand{\speedas}[3]{#1$^{+#2}_{-#3} \rm\,km \, s^{-1}$}
\newcommand{\disppm}[2]{$#1\pm#2{\rm\,km \, s^{-1}}$}
\newcommand{\hubunits}[1]{#1 $\rm \! km \, s^{-1} \, Mpc^{-1}$}
\newcommand{\Ho}{H$_0$}
\newcommand{\abmagn}[2]{$\rm M_{#1}\!\!=\!\!#2$}
\newcommand{\magn}[2]{$\rm #1\!=\!#2$}
\newcommand{\dismod}[2]{$\rm (m\!-\!M)_{#1}\!=\!#2$}
\newcommand{\absorb}[1]{A$\rm _#1$}
\newcommand{\absorbval}[2]{$\rm A_#1 \!=\! #2 $}
\newcommand{\about}[1]{$\sim \,$#1}
\newcommand{\Sn}{$S_N$}
\newcommand{\met}[1]{$\rm [Fe/H]\!=\!#1$}
\newcommand{\meterr}[2]{$\rm [Fe/H]\!=\!#1\pm#2$}
\newcommand{\meanmet}[1]{$\rm \langle [Fe/H]\rangle\!=\!#1$}
\newcommand{\metrange}[2]{$\rm #1 \! < \! [Fe/H]\! < \! #2$}
\newcommand{\valpm}[2]{#1$\pm$#2}
\newcommand{\colorval}[3]{$\rm #1-#2\!=\!#3$}
\newcommand{\color}[2]{$\rm #1\!-\!#2$}
\newcommand{\colorrange}[4]{$\rm #3 \! < \! #1-#2\! < \! #4$}
\newcommand{\excessval}[3]{$\rm E_{(#1\!-\!#2)}\!=\!#3$}
\newcommand{\excess}[2]{$\rm E_{(#1\!-\!#2)}$}
\newcommand{\app}{$\!{\rm \AA/pixel}$}
\newcommand{\Halpha}{H$\alpha$}
\newcommand{\Hbeta}{H$\beta$}
\newcommand{\Hgamma}{H$\gamma$}
\newcommand{\Hdelta}{H$\delta$}
\newcommand{\Mdot}[1]{$\rm #1 \, M_{\odot} \, yr^{-1}$}
\newcommand{\Mdotpm}[2]{$\rm #1 \, \pm #2 \, M_{\odot} \, yr^{-1}$}
\newcommand{\Mdotas}[3]{$\rm #1^{+#2}_{-#3} \, M_{\odot} \, yr^{-1}$}
\newcommand{\agerangeM}[2]{$\rm #1\!-\!#2 \, Myr$}
\newcommand{\agerangeG}[2]{$\rm #1\!-\!#2 \, Gyr$}
\newcommand{\Mrange}[3]{$\rm #1\!-\!#2 \, M_{\odot} \,\,$}
\newcommand{\seqorder}[3]{#1$<$#2$<$#3}
\newcommand{\pgcr}{projected galactocentric radius}
\newcommand{\gcr}{galactocentric radius}
\newcommand{\pgcri}{projected galactocentric radii}
\newcommand{\gcri}{galactocentric radii}
\newcommand{\gc}{globular cluster}
\newcommand{\gcs}{globular clusters}
\newcommand{\Gcs}{Globular clusters}
\newcommand{\pgcc}{proto-globular cluster candidate}
\newcommand{\pgccs}{proto-globular cluster candidates}
\newcommand{\GCS}{globular cluster system}
\newcommand{\GCSs}{globular cluster systems}
\newcommand{\MW}{Milky Way}
\newcommand{\mean}[1]{\langle{#1}\rangle}
\newcommand{\SN}{signal-to-noise}

\newcommand{\datasec}{Section 2}
\newcommand{\metsec}{Section 3}
\newcommand{\linesec}{Section 3.1}
\newcommand{\redsec}{Section 3.2}
\newcommand{\agesec}{Section 3.3}
\newcommand{\kinsec}{Section 4}
\newcommand{\rotsec}{Section 4.1}
\newcommand{\masssec}{Section 4.2}
\newcommand{\finalsec}{Section 5}

\title{Spectroscopy of Globular Clusters in M81$^1$}

\author{Linda L. Schroder}
\affil{Lick Observatory, University of California, Santa Cruz, CA 95064}
\affil{Electronic mail: linda@ucolick.org}

\author{Jean P. Brodie}
\affil{Lick Observatory, University of California, Santa Cruz, CA 95064}
\affil{Electronic mail: brodie@lick.ucsc.edu}

\author{Markus Kissler-Patig}
\affil{European Southern Observatory, Karl-Schwarzschild-Str.2 
                 85748 Garching, Germany} 
\affil{Electronic mail: mkissler@eso.org}

\author{John P. Huchra}
\affil{Harvard-Smithsonian Center for Astrophysics, 60 Garden St, Cambridge MA 01238}
\affil{Electronic mail: huchra@cfa.harvard.edu}

\author{Andrew C. Phillips}
\affil{Lick Observatory, University of California, Santa Cruz, CA 95064}
\affil{Electronic mail: phillips@ucolick.org}

\altaffiltext{1}{ Data presented herein were obtained at the W.M. Keck
Observatory, which is operated as a scientific partnership among the
California Institute of Technology, the University of California and the
National Aeronautics and Space Administration.  The Observatory was made
possible by the generous financial support of the W.M. Keck Foundation.
}

\begin{abstract}
We present moderate-resolution spectroscopy of \gcs~around the Sa/Sb
spiral galaxy M81 (NGC 3031).  Sixteen candidate clusters were observed
with the Low Resolution Imaging Spectrograph on the Keck I telescope.
All are confirmed as {\it bona fide} \gcs, although one of the clusters
appears to have been undergoing a transient event during our
observations.  In general, the M81 \GCS~is found to be very similar to
the \MW~and M31 systems, both chemically and kinematically.  A kinematic
analysis of the velocities of 44 M81 \gcs, (the 16 presented here and 28
from previous work) strongly suggests that the red, metal-rich clusters
are rotating in the same sense as the gas in the disk of M81.  The blue,
metal-poor clusters have halo-like kinematics, showing no evidence for
rotation.  The kinematics of clusters whose \pgcri~lie between 4 and 8
\kpc~suggest that they are rotating much more than those with
\pgcri~outside these bounds.  We suggest that these rotating,
intermediate-distance clusters are analogous to the kinematic
sub-population in the metal-rich, disk \gcs~observed in the \MW~and we
present evidence for the existence of a similar sub-population in the
metal-rich clusters of M31.
With one exception, all of the M81 clusters in our sample have ages that
are consistent with \MW~and M31 \gcs.  One cluster may be as young as a
few Gyrs.  The correlations between absorption-line indices established
for \MW~and M31 \gcs~also hold in the M81 cluster system, at least at
the upper end of the metallicity distribution (which our sample probes).
On the whole, the mean metallicity of the M81 \GCS~is similar to the
metallicity of the \MW~and M31 \GCSs.  To within a factor of two, the
projected mass of M81 is similar to the masses of the \MW~and M31.
Its mass profile indicates the presence of a dark matter halo.
  
\end{abstract}

\keywords{galaxies: spiral, galaxies: individual (M81,NGC3031,M31), globular clusters: general}
\section{Introduction}
\Gcs~are ideal probes of the dynamical properties and
chemical histories of their host galaxies.  They are bright enough to be
observed at large distances, old enough to have been ``witnesses'' to
the processes by which galaxies form and evolve, and are found in all
types of galaxies.  For all but the nearest galaxies, \gcs~are the sole
providers of information about the stellar population in galaxy halos.

Chemically and kinematically distinct sub-systems have been
established in the \GCSs~of both the Milky Way and
our nearest spiral neighbor, M31.  The presence of such sub-systems
places significant constraints on spiral galaxy formation theories and
may also bear on understanding the formation and evolution of
elliptical galaxies.  It has been suggested \citep{schweizer86}
that some elliptical galaxies may form via the merging of spirals.
In a gaseous merger, the resulting \GCS~would
be expected to contain any \gcs~originally
associated with the merger participants and perhaps new clusters formed
during the merger \citep{AZ92}.  While study of \GCSs~around
ellipticals is important for validating or refuting the merger
hypothesis, interpreting such studies requires understanding the
natural variations in the observed properties of \GCSs~around spirals.
Key properties such as age and metallicity are most reliably
constrained through spectroscopy, while kinematic and individual
element abundance information are obtained exclusively so.  The advent
of 10-meter class telescopes and multi-object spectrographs
has made progress on this front feasible.  

The Milky Way \GCS~is naturally the benchmark by which the systems of other
spirals are measured.  One of its most striking features is its
bimodal metallicity distribution, with peaks at \met{-1.6}~and
$-0.5$ \citep{zinn85,AZ88,armandroff89}.  Furthermore, these metallicity
sub-populations are spatially and kinematically distinct
\citep{zinn85,hesser_etal86,AZ88,armandroff89}.
The metal-rich clusters comprise a centrally-concentrated,
flattened, rotating system with a low velocity dispersion, 
referred to by \citet{zinn85} as the disk population.
\citet{armandroff89} associated these disk clusters with the Milky Way's
thick-disk component, based on similar spatial, kinematic and chemical
properties of the two systems.  More recently, it has been reported that
the properties of the metal-rich clusters lying within 3\kpc~of the
Galactic center have more in common with the bulge than with the thick
disk \citep{minniti95,cote99}.  \citet{burkert_smith97} further explored
sub-structures in the metal-rich \gc~population, identifying a group
belonging to the galactic bar and one comprising a ring of
rapidly-rotating clusters with a low velocity dispersion and
galactocentric radii between 4 and 6 \kpc~(these are the thick-disk
clusters of Armandroff (1989)).  To first order,
\citet{zinn85} found the properties of the Milky Way's metal-poor \gcs~to
be well matched with those of the halo field stars:
they are spherically distributed, rotate slowly and have a large
line-of-sight velocity dispersion.  However, \citet{zinn93} reported 
substructure in the Milky Way's halo clusters as well.

The M31 \GCS~was the first system around another spiral to be studied in
detail.  A number of imaging programs resulted in several catalogues and
refined lists of \gc~candidates
\citep{crampton85,battistini80,battistini87,battistini93,reed_etal95}.  Insight
into the chemical and kinematic nature of M31's \GCS~resulted from the
optical spectroscopy studies of \citet{HSV82}, \citet{KHS89},
\cite{HBK91} and \cite{BH91}.  All came to the same conclusion:  the
\GCS~of M31 is, for the most part, very similar to the Milky Way system.
\cite{HBK91} derived metallicities for 149 clusters and found
them to be consistent with the metallicity distribution of the
\MW~\GCS, but they were unable to distinguish any obvious bimodality.  They
noted that large uncertainties in their individual metallicity
measurements could be obscuring its presence.  \citet{Ashman_Bird_93}
subsequently performed a statistical analysis on the \citet{HBK91} data
and found quantitative evidence for a bimodal metallicity distribution
at the 98.4\% confidence level.  New optical and infrared
color distributions by \citet{barmby} definitively support the 
presence of bimodality.

Based on velocities of 60 clusters, \citet{HSV82} detected a modest, but
significant rotation for the entire system.  Subsequent studies have yielded
evidence for kinematic sub-systems as well.   \citet{HBK91} reported
that the metal-rich clusters in the inner few \kpc~of M31 are rapidly
rotating, while at larger radii they observed little rotation in the M31
\gcs, regardless of cluster metallicity.  \cite{barmby} recently repeated
the kinematic analysis of \cite{HBK91} with an additional 74 clusters
and derived the same result.

As similar to the \MW~system as M31's \GCS~appears, there do seem to be
some genuine differences.  The M31 clusters contain, on average, a higher abundance
of nitrogen than Milky Way \gcs.  This result was first reported as a CN
enhancement in the spectroscopic studies of \citet{burstein_a} and
\citet{BH91} and more recently in the ultraviolet spectroscopy of
\citet{ponder98}, who established it as a nitrogen overabundance.
\citet{BH91} also reported evidence of calcium enhancement in the M31
cluster spectra.  In addition, the M31 system contains a handful
\gcs~which appear to be ``young'', based on the strength of their
Balmer-line absorption
\citep{sargent77,cowley88,elson_walterbos88,BH91,barmby}.  No such
clusters have been observed in the Milky Way.

Globular cluster systems around a few other spirals have been examined
to a more limited extent.  Photometric
\citep{christian82,cohen84,sharov_lyutyj84,christian88,chandar_a,chandar_b,chandar_c}
and spectroscopic \citep{schommer91,BH91} studies of M33, the late-type,
low-luminosity, Local-Group spiral, have revealed that it contains a
number of young and intermediate-age compact star clusters.  In
addition, it contains about 25 typical old \gcs~in a halo distribution
with little net rotation.

Beyond the Local Group, M104 and M81 are the only spiral galaxies
whose \GCSs~have been studied spectroscopically.  Velocities for 34
M104 \gcs~were obtained by \citet{bridges97}, who found
marginal evidence for rotation.  The quality of their data was too poor
to measure metallicities for individual clusters.  They produced a rough
estimate of the mean metallicity of the system from a composite spectrum
of all 34 objects, finding \meanmet{-0.7}$\pm{0.3}$.  This value is
substantially higher than the metallicity of the \MW~and M31 globular
cluster systems, but M104
is extremely bulge-dominated.  It is also significantly more luminous than the
\MW~or M31, and its mean \gc~metallicity is similar to the
values found for \gcs~around elliptical and giant elliptical galaxies
with a luminosity similar to that of M104 (see Table 4 in Bridges
\etal~1994.)

At a distance of 3.7 Mpc \citep{freedman94} M81 is at roughly half the
distance of M104.  It is an Sa/Sb-type spiral, very similar to M31, and
roughly as massive as the Milky Way.  Its proximity and morphology
make it a good candidate for reaching beyond Local Group toward
a detailed, high \SN~spectroscopic study of a spiral galaxy \GCS~for
comparison to the \MW~and M31 systems.  In their
study of extragalactic \gcs, \citet{BH91} derived
spectroscopic metallicities for eight clusters in M81 and found a sample
mean of \meterr{-1.46}{0.31}.  \citet{PR95} conducted a
photometric study of the M81 \GCS~and
compiled a catalogue of \gc~candidates, using proper
motion to minimize contamination from foreground stars.  In a
spectroscopic follow-up, \citet{PBH95} obtained low
\SN~spectra of 82 candidates, 25 of which were established as {\it
bona fide} members of the M81 \GCS.  The generally poor \SN~of their
data resulted in rather large uncertainties in the metallicities of
individual clusters.  They derived a more precise mean metallicity of their
sample from a composite spectrum, finding \meanmet{-1.48}$\pm{0.19}$.
Their analysis of the cluster velocities showed a hint of rotation in
the M81 \GCS.

In this work we extend the spectroscopic studies of \citet{BH91} and
\citet{PBH95} with significantly higher \SN~spectra of an additional 16
members of the M81 \GCS.  \datasec~describes our observations and data
reduction procedure.  A discussion of the abundance ratios, mean
metallicities and ages of the M81 clusters and a comparison of these
properties to M31 and the \MW~\gcs~can be found in \metsec.  In
\kinsec, we combine our velocity data with the velocities of additional
clusters from the work of \citet{PBH95} and \citet{BH91}, and discuss
the kinematics of the resulting sample.  \finalsec~contains a summary of
our results and a brief discussion of their implications for using
\GCSs~as probes for studying galaxy formation and evolution.

\section{Observations and Data Reduction}
\subsection{Observations}
The targets were selected from the candidate list of \citet{PR95}.
The specific candidates observed were selected on the basis of
brightness and the desire to simultaneously observe the maximum
possible number of candidates with the spectrograph in multi-object
mode.  Spectra of 16 \gc~candidates were obtained on 1994 November 30
using the Low Resolution Imaging Spectrograph (LRIS)
\citep{oke95} on the Keck I 10-m telescope in $\sim\!0.$\arcsec8 seeing.
The candidates were observed through a single slit mask.  Weather
and technical problems prevented observations through two additional
masks, so our sample is spatially biased toward one side of the galaxy.
The 600 l/mm grating, blazed at 7500\AA, and slitlets of
1\arcsec~width resulted in a dispersion of 1.25~\app~and a spectral
resolution of 5-6\AA.  Because the horizontal positions of the
slitlets varied across the slit mask, the spectral range falling on
the detector was different for each slitlet.   The shortest wavelength
observed was 3700\AA~and the longest was 7420\AA.   Each candidate
spectrum covers 2560\AA~between those two extremes.  The region of
common overlap for all candidates is 4680-6260\AA.   We obtained 5
$\times\, 900\,{\rm s}$ integrations for a total integration of
$4500\,{\rm s}$.  A complete list of the observed cluster candidates'
positions, magnitudes and colors is provided in \mytab{posntable}.
\myfig{posfig} shows the positions of the candidates with respect to the
galaxy center, along with the positions of established clusters from the
work of \citet{PBH95} and \citet{BH91}.

\subsection{Data Reduction}
These data were reduced using the following procedure.  A mean bias,
assumed to be the mean level of the CCD overscan region, was
subtracted from both science and calibration images.  In addition, the
small-scale bias structure was determined by combining multiple
one-second dark exposures.  This bias structure was subtracted from
all other science and calibration images.  After the images were
cleaned of cosmic rays, a map of the small scale variations in
detector sensitivity and grating efficiency was removed by fitting a
fifth- or sixth-order polynomial to a halogen lamp spectrum taken
through each slitlet in the mask and dividing the lamp spectra by the
polynomial.  The resulting normalized images were divided into the
science images to remove the small scale variations.

Each slitlet was subsequently processed using the IRAF\footnote{\small
IRAF is distributed by the National Optical Astronomy Observatories,
which are operated by the Association of Universities for Research in
Astronomy, Inc., under cooperative agreement with the National Science
Foundation.} routines of the {\tt noao.twodspec} package.  {\tt Ident},
{\tt reident} and {\tt fitcoords} were run on spectra of Hg, Ne, Ar and
Kr lamps to determine the wavelength solution and then {\tt transform}
was used to apply the appropriate solution to the two-dimensional
spectra of the candidate objects.  The object spectra were extracted
using the IRAF routines of the {\tt noao.apextract} package, employing
the optimal spectral extraction scheme detailed by \citet{horne86}.  The
polynomial used to approximate the background was typically a linear
fit.  For a few candidates, however, proximity to end of the slitlet
required approximating the background as a constant.
  
The extracted candidate spectra were flux-calibrated using a long
slit spectrum of the standard star EG145 taken on the same night as
the candidate observations.  Since its location in the dispersion
direction on the mask determines what wavelength region falls onto the
detector for an individual slitlet, the wavelength coverage for some
candidate cluster spectra is not coincident with the coverage for the
flux standard.  The result is that the continuum shape is only
accurately determined in the spectral region of 3900\AA-6500\AA.  None
of the absorption features of interest lie beyond 6500\AA, while two
spectral features of interest (see \metsec) require measurements below
3900\AA.  The final spectra of all 16 candidates are shown in
\myfig{specfig}. They have been corrected for foreground reddening in the
direction of M81 using the IRAF task {\tt noao.onedspec.deredden} with
\excessval{B}{V}{0.1} and ${\rm R_V = 3.1}$ \citep{burstein_b}, then
shifted to the rest frame.

M81's low systemic heliocentric velocity (\speed{-34}) precludes the
establishment of \gc~candidates as {\it bona fide} M81 \gcs~on the basis
of radial velocity alone.  \citet{PBH95} circumvented this problem by
comparing each candidate's color to the relative strengths of the
calcium K and \Hdelta~absorption lines in its spectrum.  Unfortunately,
the low throughput of LRIS below 4000\AA~and the non-coverage of that
part of the spectrum for many of our candidates prevented us from doing
likewise.  However, since our field is within 5\arcmin~of the center of
M81 where the cluster surface density is high, it is very likely that
candidates whose spectra contain typical globular cluster features are
members of the M81 \GCS.  This assertion is supported by the fact that
the \citet{bahcall_soneira_80} models predict fewer than 2 galactic
stars in our 48 arcmin$^2$ field (the LRIS field is 6\arcmin
$\times$ 8\arcmin) in the direction of M81 and in the color and
magnitude range of our candidates.

Fifteen of the 16 candidates observed have spectra that are typical of
\gcs~and we henceforth treat them as such.  Only one, object 12,
appears at first to be ruled out as a \gc~on the basis of its spectrum
(see \myfig{specfig}).  The sharp upturn at the blue end of the spectrum
and what appears to be broad-line Balmer emission are decidedly
``un-globular''.  Still, there are genuine globular cluster features in the
spectrum of this object, occurring at wavelengths that are consistent
with it belonging to M81.  Curiously, the photometric study of
\cite{PR95} cataloged a \color{B}{V} value of 1.0 for this object, which
is well within the normal color range for a \gc.  The color of an object
with a spectrum such as object 12 would certainly be much bluer, and likely
would not have satisfied the \citet{PR95} color criteria for inclusion
in their catalogue of candidates.  Thus we suspect that a
transient phenomenon was occurring in the cluster at the time of our
observation.  Preliminary reductions of a spectrum of object 12
recently obtained with the Echellette Spectrograph and Imager (ESI) on
the Keck II telescope has confirmed our suspicion.  Object 12 no longer shows 
an upturn at the blue end of the spectrum and no broad-line emission
features are present.  Thus we consider object 12 to be a genuine M81
globular cluster and include it in our kinematic analysis of the M81
\GCS~(\kinsec).  However, because the continuum shape is pivotal for
accurate measurements of absorption line strengths and hence
metallicity, we exclude object 12 from our chemical abundance and
metallicity analysis in \metsec.  The exact nature of the phenomenon
occurring during the original observation of object 12 will be the subject of a
forthcoming paper.

\subsection{Measurement of Radial Velocities}
A radial velocity was determined for each cluster by
cross-correlating its non-flux-calibrated spectrum with three
high signal-to-noise templates of M31 \gcs.
The actual cross-correlation was accomplished using the
IRAF routine {\tt fxcor} in the {\tt rv} package.
One of the templates (225-280) was obtained on the same run as these
data and the other two (225-280 and 158-213) were obtained on a
subsequent run, also with LRIS and at the same resolution.  The ID
numbers and heliocentric velocities of the template objects are from
\citet{HBK91}.  The template spectra have the same resolution as the
cluster spectra, 5-6\AA, corresponding to a velocity resolution of
roughly \speed{300}.  The results from the three templates agree quite well,
the differences being a small ($\sim$5-10\%) fraction of the velocity
uncertainties returned by {\tt fxcor}.  The heliocentric
line-of-sight velocities presented in \mytab{veltable}~are the average
of the results from the three templates.

\section{Abundances and Metallicities}
\subsection{Metallicities}
We derived each cluster's mean metallicity by taking the error-weighted
average of metallicities predicted by the strength of several absorption
features.  

The absorption line indices were measured as described in
\citet{BH90}, but our method for estimating the uncertainties was
handled slightly differently because our background subtraction
technique was different.

The extraction routines we used (see section
2.2) allow for the output of a ``variance'' spectrum, which is actually
the standard deviation for each pixel in the extracted,
background-subtracted spectrum, based on
the quality of the polynomial fit to the background and the number of
rows over which the spectrum was summed before extraction.  We used this
``variance'' spectrum to estimate the photon uncertainty on the
average flux per angstrom in a given bandpass by summing the
``variance'' pixels in quadrature over entire bandpass and then dividing
by the bandpass width:

\begin{center}$\langle\sigma_{BP}\rangle={ \left[\sum\limits_{j=1}^n \sigma_{pix}^2\right]^{1\over 2}\over {\lambda_2 - \lambda_1} } $, 
\end{center}

where $\langle\sigma_{BP}\rangle$ is the photon uncertainty on the average
flux per angstrom in a generic bandpass, $\sigma_{pix}$ is the standard
deviation in a given pixel (as given by the ``variance'' spectrum), $n$
is the number of pixels in the bandpass and $\lambda_2$ and $\lambda_1$
are the lower and upper wavelength limits of the bandpass, respectively.

As in \citet{BH90}, we computed the uncertainty in the measured value of
an index in a particular spectrum, $\sigma_i$, by summing in quadrature
the photon uncertainties on the flux of each bandpass used to compute the
index, including the absorption feature bandpass itself and the two
continuum bandpasses:

\begin{center}$\sigma_i^2 =\sigma_{C1}^2 + \sigma_{L}^2 + \sigma_{C2}^2$,
\end{center}

$C1$ and $C2$ stand for the two continuum bandpasses and $L$ stands for
the absorption line bandpass.

For each cluster, both the indices themselves and the photon errors on
the indices were computed separately for each of our five
integrations.  The final index values are a weighted average
of the indices derived from each
integration using ${1\over\sigma_{i}^2}$ as the weight.  The final index
photon errors, $\sigma_I$, are given by the reduced photon error:

\begin{center}$\sigma_I=\left({1\over \sum\limits_{k=1}^N\sigma_i^2}\right)^{-{1 \over 2}}$.
\end{center}

\mytab{BHmettable}~gives the
metallicity predicted by each spectral feature and the error-weighted
mean metallicity of each cluster.  The uncertainty in the metallicities
predicted by a given individual index were derived by simply propagating
the error in that index, $\sigma_I$, through the linear relationships
between the index and metallicity derived in \citet{BH90}.  The
uncertainty in the error-weighted mean metallicity for each cluster is,
in a sense, an internal error since it is the standard of the mean.
The individual Brodie \& Huchra index
values may be found in \mytab{BHindtable}.  Two of the features used by
\citet{BH90} to predict metallicity (CNB and Delta) require measurements
below 3900\AA.   Since, as mentioned in \datasec, the continuum shape of
the spectra is uncertain for regions below 3900\AA~and above 6500\AA~we
measured the index value and predicted metallicities for these two
features (when the spectral coverage was sufficient) but excluded them
from our derivation of the clusters' mean metallicities.


The metallicities of our sample clusters span a range of
\metrange{-1.49}{-0.21} with an error-weighted mean of
\meanmet{-0.91}$\pm{0.09}$ and a dispersion of \valpm{0.29}{0.07} dex.
This range is narrower than that spanned by either the \MW~or M31
\GCSs.  This may be due to a bias in our observed sample.
As shown in \myfig{colorhistfig} the colors of our sample clusters
are biased toward the red end of the color distribution.
The mean \pgcr~of our sample is $2.6\kpc$. If a cluster population
associated with a metal-rich bulge or thick disk exists in M81, then our
luminosity-selected sample taken from a $6\arcmin \times 8\arcmin$
region centered at $2.6\kpc$ is likely dominated by such clusters,
and the mean metallicity of our observed sample is not likely to
accurately represent the mean metallicity of the entire M81 cluster system.
Indeed, the mean \pgcr~of the \citet{PBH95} sample is
significantly larger (9.6 $\kpc$) and yields a rather lower mean
metallicity value of \meterr{-1.48}{0.19} as measured from a composite
spectrum.  \citet{BH91} derived nearly the same result.  To within the
errors, this value is consistent with the peak of the metal-poor
population in the metallicity distributions of the Milky Way and M31
cluster systems ($-1.5$ and $-1.6$, respectively).
The mean metallicity of the M81 Keck sample is somewhat lower than the mean of
the metal-rich peak of M31 and \MW~cluster metallicity distributions.
With such a small sample size we cannot say whether the metal-rich M81
clusters, as a group, have a lower metallicity than their counterparts
in the other systems or if we are simply seeing the inclusion of 
a few clusters which belong to a metal-poor halo but have small
projected galactocentric radii.

The absence of clusters in our sample with [Fe/H] greater than $-0.2$ is not
really surprising, given the rarity of such clusters in the \MW~and M31.
Only 7 of the 150 M31 clusters studied by \citet{HBK91} have
metallicities higher than $-0.2$, and just three of the 134 \MW~clusters
listed in the McMaster's catalog \citep{harris96} fit this criteria.

In an attempt to estimate the mean metallicity of the entire M81 \GCS~we
derived the arithmetic mean of all the M81 clusters for which
spectroscopic metallicities have been established.  An error-weighted
mean would not be useful here because the metallicity errors for the
Keck sample are much smaller than the errors in the other two studies
and the clusters from the Keck sample would entirely dominate the
result.  In Table 3 of their paper, \citet{PBH95} indicate that three of
the clusters in their study overlap with the \cite{BH91} sample.
After comparing the coordinates and colors of objects in the two studies
we have identified all of the objects from \citet{BH91} as candidates
in the \citet{PR95} catalogue (see \mytab{johntable}).  There are indeed three
clusters which are present in both studies but one of them was
misidentified by \citet{PBH95}.  Using the nomenclature of their Table
4, it is cluster Id50415 that should be matched with G19, not Is50401.
At any rate, we adopt the error-weighted average of the metallicities
derived in the two studies for these three clusters.  The arithmetic
mean metallicity of the resulting sample of 44 clusters is
$\mean{{\rm[Fe/H]}}=-1.25\pm0.13$ where the error is the standard error
of the mean.    The mean value is entirely consistent with simply taking
an arithmetic mean of metallicities of either the Milky Way
($-1.30\pm0.05$ dex) or M31 ($-1.14\pm0.04$ dex) \GCSs.   The dispersion about
the mean is $0.88\pm0.10$ dex, larger than for the other two systems
(\about{0.55} dex).  This is likely due to the sizeable uncertainties in
the metallicities from \citet{PBH95} and \citet{BH91} rather than an
intrinsically larger spread in the metallicities of the M81 \gcs.

\subsection{Internal Reddening in M81}

Using the \color{B}{V} color-metallicity relation established for
\MW~\gcs~by \citet{couture90}, we computed metallicities predicted by
their \color{B}{V} colors for the clusters in the Keck sample.  Object 
12 was excluded from this analysis (see \datasec).  \myfig{comparemetfig} shows the
photometric metallicities plotted against the metallicities derived from
their spectra as described in \linesec.  For reference we also show
\MW~clusters for which foreground reddening is low
(\excess{B}{V}$<0.4$) and the line where clusters would lie if
photometric and spectroscopic metallicities were in perfect agreement
In the upper panel, the M81
cluster colors were corrected for foreground reddening only.  Even after
such a correction, the majority still lie above the line, i.e., their
spectroscopic metallicities are lower than their colors would predict.
An obvious explanation is that the light from the M81 clusters suffers
internal reddening, which is likely to be variable across the disk of
the galaxy.  The range of color excess that would have to be applied to
get all of the M81 clusters above the line to lie directly on it is
$0.01 <$ \excess{B}{V}$< 0.24$.  Given both the intrinsic
scatter and non-linearity of the true color-metallicity relation,
we cannot know the precise reddening for the individual clusters.
However, the mean of the range quoted above {$\rm \langle
E_{(B\!-\!V)}\rangle\!=\!0.12$} is likely a reasonable estimate of the
{\it average} internal reddening in M81.
This value is slightly lower
than the value quoted by \citet{PR95} but to within the errors they are
consistent.  The lower panel of \myfig{comparemetfig} shows the results
of applying an additional 0.1 mag correction to the cluster colors and
recomputing the photometric metallicities.

To investigate the degree to which reddening could affect our derived
cluster metallicities, we further de-reddened the spectra of the
three clusters which required the largest corrections to bring their
photometric and spectroscopic metallicities into agreement and
remeasured their metallicities.  We found that de-reddening the
spectra systematically increases our derived mean metallicities, but
the average change is only 0.05 dex, which is a fraction of the
average statistical uncertainty of \about{0.20} dex.  The errors
themselves also systematically increase, due to increased scatter in
the metallicity values predicted by individual spectral features.
This exercise demonstrates that spectroscopic metallicity
determinations are far less vulnerable to errors introduced by
internal reddening, a significant problem for the method of
determining metallicities by photometry alone.  Spectroscopy is
especially advantageous when observing clusters in the disks of spiral
galaxies where patchy dust can make the internal reddening highly
variable. 

\subsection{Individual Absorption Line Strengths}
Our spectra have a sufficiently high \SN~ratio to compare the strengths of
individual absorption features to those of \MW~and M31 \gcs.
The complete suite of Lick/IDS indices \citep{gonzales93,trager98}
for the \gcs~in our M81 sample are presented in \mytab{LIDStablea} and
\mytab{LIDStableb}.
\myfig{elementfig} shows the strengths of G4300, NaD, TiO1, CN2,
$\mean{{\rm Fe}}$ and Mg2 plotted against [MgFe] strength for M81,
\MW~and M31 \gcs.  \citet{gonzales93} defines $\mean{{\rm
Fe}}$ as the arithmetic mean of the Fe5335 and Fe5270 indices, and [MgFe] as
${\rm \sqrt{\mean{Fe} \times Mgb} }$.  The relations between these
indices and the [MgFe] index for the M81 clusters in the Keck
sample are generally the same as for \MW~and
M31 \gcs.  We had sufficient spectral coverage to measure the
CN2 strengths for five M81 clusters, in which we see no evidence of a CN
enhancement (see panel {\it d} of \myfig{elementfig}).

The Mg2 index appears to be slightly depressed in M81 clusters when
compared to M31 and \MW~clusters with similar [MgFe] values (see panel
{\it e} of \myfig{elementfig}).  In \myfig{ironfig} the values of
Fe5270, Fe5335 and Fe5406 are plotted against Mg1, Mg2 and Mgb for M81,
M31 and \MW~clusters.  Mg2 appears depressed in M81's \gcs, relative
to Fe5335 but it is completely in line for the other two iron indices.  The
depression of Mg2 in \myfig{elementfig} can be better described as an
excess of Fe5335, causing enhanced [MgFe] values.  But is it a real
effect?  We do not think it is likely.  The strengths of the Fe5335 and
Fe5270 indices, whose average makes up $\rm \mean{Fe}$ and thus affect
the value of [MgFe], can be influenced by contamination from other
elements.  Of all the Lick/IDS iron indices, only Fe5406 is a pure
measure of iron \citep{trager98}, and the Fe5406 index values of the
M81 clusters are quite consistent with the \MW~and M31 clusters for all
three of the magnesium indices.  The relative abundances of magnesium
and iron in M81's \gcs~appear to be completely consistent with \MW~and
M31 \gcs, at least at the upper end of the metallicity range.

\subsection{Ages}
In \myfig{agefig} we plot the age-sensitive index \Hbeta~against Mg2
for M81~\gcs.  For reference we also show these index values for
individual Milky Way and M31 \gcs, taken from the work of
\citet{trager98}.  Isochrones from the evolutionary-synthesis models of
\citet{worthey94} and from \citet{FB95} are overplotted.  Although there
are significant differences between the models (the Worthey models
give lower \Hbeta~absorption values at a given age and metallicity,
probably because these models do not include blue horizontal-branch
stars in their population-synthesis scheme) it is clear that all but
one of the M81 clusters in our sample are old, i.e., their ages are
similar to those of Milky Way and M31 \gcs.  The
exception is object 15, whose \Hbeta~index value is quite high when
compared to the other clusters in the sample. Strong \Hbeta~absorption
in \gcs~is often interpreted as indicating low cluster
metallicity, since the horizontal branches of metal-poor clusters are
typically bluer than those of metal-rich clusters.  However, the
model tracks are isochrones, showing the behavior of the
\Hbeta~equivalent width for simple stellar populations of a single
age, for a range of metallicities. The strong \Hbeta~absorption in
this cluster cannot be explained by metallicity alone.  In addition,
the color of this cluster is much bluer than would be expected on the
basis of its mean metallicity.  This object may in fact be a young (3
Gyr) \gc.  Similar clusters have been observed in both M33 and
M31 (Brodie \& Huchra, 1990; 1991).

\section{Cluster Kinematics}
Using the maximum likelihood method of \citet{pryor_meylan93} we find a mean
line-of-sight velocity for the Keck sample of \speedpm{-105}{32} and
a velocity dispersion of \speedpm{123}{23}.
The mean velocity value differs significantly from both the heliocentric systemic
velocity of M81 (\speedpm{-34}{4}) and the median cluster radial
velocity found by \citet{PBH95} of \speedpm{-18}{32}.  We note, however,
that the spatial extent of our sample is small and biased to one side
of the galaxy (see \myfig{posfig}).  The Keck sample
shows no significant evidence for rotation, but the spatial bias could
be obscuring its presence.
To increase the sample's size and spatial extent, we include the
velocities of 25 confirmed clusters from the study of \citet{PBH95} as
well as the unpublished velocities of eight additional
objects from \citet{BH91}.  As mentioned in \linesec, we have matched
all eight of the Brodie \& Huchra objects with members of the \citet{PR95} candidate list (see
\mytab{johntable}).  Based on their colors and/or proper motions, two of
them are likely foreground stars.  The remaining six objects are
confirmed M81 globular clusters, three of which were also observed by
\citet{PBH95}.  For these three we have adopted the error-weighted mean
of the velocities from the two studies.  We combined the three data sets
into a sample containing 44 M81 clusters, which we henceforth refer to
as the ``full sample''.

The full sample has a mean heliocentric line-of-sight
velocity of \speedpm{-34}{23}, in excellent agreement with M81's
systemic velocity.  The line-of-sight velocity dispersion of the
full sample is \speedpm{133}{19}, generally consistent with the velocity
dispersions of the \GCSs~of the \MW~($\sigma\sim 100$; Armandroff, 1989; 
DaCosta \& Armandroff, 1995) and M31 ($\sigma\sim 150$; Huchra, 1993;
Huchra, Brodie \& Kent, 1991).  The line-of-sight dispersion of
M104's \GCS~is significantly larger ($\sigma\sim 260$; Bridges \etal~1997),
but that is not unusual given its higher luminosity.

\subsection{Rotation}

\myfig{rot1fig} shows the positions of the M81 \gcs~marked with
circles at the location of each cluster.  The size and thickness of
the circles are indicative of the magnitude and direction of the
cluster radial velocities.  Receding clusters ({\it thick-lined circles})
tend to be located in the northwest, approaching clusters ({\it thin-lined
circles}) tend to lie more to the southeast.  This suggests that
rotation is present in the M81 \GCS~in a sense that is consistent with
the disk \citep{adler_westphal96}.  A more quantitative analysis is
demonstrated in \myfig{rotcolorfig}.  Panel {\it a} shows the
velocities (relative to the galaxy) of the full sample of clusters
as a function of position angle with respect to the galaxy center.
We used the maximum-likelihood method of \citet{pryor_meylan93} to
determine a rotation velocity and rotation axis for the sample.
The analysis assumes that the system is rotating as a solid body.
Data points taken from true solid-body rotators would
follow the sinusoids in \myfig{rotcolorfig} quite closely, although
some scatter would necessarily be present, since clusters at the
same position angle will have a variety of perpendicular distances
from the axis of rotation.  The rotation velocity quoted for each
sub-sample in \myfig{rotcolorfig}, \myfig{rotspacefig} and in
\mytab{kintable} is the linear speed of
a point in the solid body which the sinusoid represents taken at
the mean \pgcr~of the sub-sample.  For the full sample we find a
rotation velocity of \speedpm{85}{32} about an axis that is
\valpm{17}{28} degrees east of north.  This axis, shown by the arrow in
\myfig{rot1fig}, is roughly aligned with the galaxy's minor axis and
also the rotation axis for M81's gas \citep{adler_westphal96}.   For
comparison, the same analysis of 182 M31 \gc~velocities yields a
rotation amplitude of \speedpm{55}{14} and a line-of-sight dispersion
of \speedpm{138}{8}.  Here too, the rotation axis derived for the entire
sample of M31 clusters is roughly aligned with M31's minor axis.

Given the division of the Milky Way and M31 \GCSs~into
kinematic sub-systems which differ in their spatial and chemical
characteristics, it is of interest to look for evidence of such
sub-populations in the \GCS~of M81.  Since the spectroscopic
metallicities of about two-thirds of the M81 full sample have large
uncertainties, the clusters' \color{B}{R} colors must suffice for assigning
them to one metallicity population or another.   In
the \MW~\gcs, the break between the metal-rich and metal-poor
clusters corresponds to
\color{B}{R}\about{1.3}, so we adopt this value for separating the
M81 clusters into red (metal-rich) and blue (metal-poor) sub-samples
containing 33 and 11 members, respectively.  We also examined a group of
the 18 reddest clusters (\color{B}{R}$ >\!1.5$).  We realize that these
sub-samples represent only rough metallicity groups and that, because
of internal reddening (see \redsec), some clusters which belong in the
low-metallicity group could be ``scattered'' into the high-metallicity
group, i.e. the red sub-samples may contain clusters that are not truly
metal-rich.  

The results of our kinematic analysis are summarized \mytab{kintable}.
\myfig{rotcolorfig} illustrates that rotation is present in both red
sub-samples at the 2 $\sigma$ level, while the blue sub-sample shows no
evidence for rotation.  Additional analyses of sub-samples based on
spatial characteristics alone showed that it is clusters at intermediate
\pgcri~that contribute most strongly to the global rotation signature
(\myfig{rotspacefig}).  As \mytab{kintable} shows, the precise value of
the rotation velocity for intermediate-range clusters varies depending
on the choice of radius.  We found the most significant detection in the
3--7 \kpc~sub-sample (\speedpm{157}{37}), but the other reasonable
choices of radius limits (2--6 \kpc~and 4--8 kpc) yielded strong
detections also.  The presence of strong rotation in M81's \gcs~at
intermediate \pgcri~invites a comparison to the \MW~thick-disk clusters
noted by \citet{armandroff89} which were further described by
\citet{burkert_smith97} as a ``ring'' of rapidly rotating metal-rich
\MW~clusters in a 2 \kpc~wide annulus at intermediate \gcr.
Most of the M81 clusters in the intermediate-radius sample are members
of the red, metal-rich sample, i.e., most have $\rm B-R>1.3$.
Unfortunately our sample is not large enough to make a direct comparison
between metal-rich and metal-poor clusters in specific radius ranges,
but at the very least our data are consistent with M81's metal-rich
clusters at intermediate projected radii being associated with a thick
disk in M81.
  
One may wonder why such a population was not detected in the \GCS~of
M31, given the much larger sample of clusters for which precise
velocities and reliable spectroscopic metallicities have been measured.
The main result of \citet{HBK91}'s kinematic study was that the inner
metal-rich clusters in M31 were the most rapidly rotating.  The analysis that 
\citet{Ashman_Bird_93} performed on the \citet{HBK91} data, showed
strong statistical support for the division of M31's \GCS~into disk and halo
components.  The method employed by \citet{HBK91} was quite different than
the one we use here.  They used each cluster's projected distance along the 
galaxy major axis to bin the metal-rich and metal-poor clusters into
spatial sub-samples.  Then, for each metallicity group they simply
computed the difference in the {\it average} heliocentric radial
velocity for clusters on either side of the galaxy, lying beyond a given
distance along the major axis.  They estimated the rotation velocity to
be one-half of this value.  They found the difference between the
metal-rich and metal-poor cluster rotation velocities to be most
pronounced at very small distances along the galaxy major axis ($<1.5$
\kpc) with less pronounced differences persisting out to distances of
roughly 8 \kpc.  They did {\it not} examine sub-samples in specific
ranges of projected radii along the galaxy major axis, so their analysis
did not address the question of whether the rotation of clusters at
intermediate projected radii might show even stronger differences
between the metal-rich and metal-poor groups.

We performed the maximum-likelihood kinematic analysis employed for the
M81 clusters on 166 M31 clusters for which spectroscopic metallicities
exist \citep{barmby}.  \mytab{m31table} shows that, as in the Milky Way,
the {\it most} significant difference between the rotation of the
metal-rich and metal-poor clusters occurs at intermediate \pgcri.  For
all reasonable choices of radii, the difference between the metal-rich
and metal-poor clusters is quite striking.  Admittedly, treating the
rotation axis as a free parameter greatly increases the chance for
detecting rotation in a small sample.  As a case in point, our analysis
yields a weak detection of rotation with approximately the same
amplitude but with very different rotation axes for metal-rich and
metal-poor clusters lying inside 2 \kpc.  However, most of the
metal-poor sub-samples have poorly-constrained rotation axes and have
rotation-velocity-to-velocity-dispersion ratios that are consistent with
their being supported primarily by thermal motion rather than by
rotation.  (The metal-rich M31 clusters inside 2~\kpc~also have a low
${v_{rot}\over\sigma_{los}}$, indicating that they may be associated
with M31's bulge.) The metal-rich sub-samples all have their rotation
axes roughly aligned with the galaxy's minor axis and they are rotating
in the same sense as M31's gas \citep{loinard_etal95}.  Still, as a
check on the extent to which treating the rotation axis as a free
parameter could influence the value of the rotation velocity derived for
the M31 sub-samples, we fixed the rotation axis to be aligned with M31's
minor axis and repeated the analysis.  To within the uncertainties, this
procedure little or no change in the rotation velocities for all
sub-samples except for the metal-poor clusters inside 2 \kpc, which
exhibit no rotation at all about the galaxy's minor axis.  Thus the main
result of \citet{HBK91}, who assumed the clusters were rotating about
the galaxy minor axis, is recovered here.  However, our analysis also
reveals a potential thick-disk population among M31's metal-rich \gcs,
analogous to those found in the \MW~and for which we see strong evidence
in the \gcs~of M81.

While nature has not provided a sufficient number of
metal-rich clusters residing at intermediate \pgcri~to ever construct
samples large enough to allay concerns about small-number statistics,
the fact that all three spiral galaxies whose \GCSs~have been
studied in reasonable spectroscopic detail show evidence for similar
populations offers encouragement that they are part of analogous stellar
populations, which are possibly the thick disk components of these
galaxies.

\subsection{Projected Mass}
The velocity distribution of the \gcs~can be used to give
an estimate of the mass of the parent galaxy.  To this end we follow
the method of \citet{heisler_etal85} and use the projected mass estimator 
\centerline{$M_p = {f_p \over G(N-\alpha)}(\sum_i^N V^2_i r_i)$.}
We adopt a value of $\alpha=1.5$, as recommended by \citet{heisler_etal85} and
assume a value of $f_p = {32 \over \pi}$.  The latter corresponds to the
case that the clusters are on isotropic orbits.  If the
clusters are on radial orbits, then the derived projected mass will be
lower by a factor of 2. Based on the full sample of 44 clusters, the
projected mass of M81 enclosed within $21.3\kpc$ is $\rm
4.0\pm0.8\times10^{11}\msun$.  The error quoted is the
dispersion in the distribution of projected masses computed from 1000
simulated data sets generated using the bootstrap method.  This is
only an approximation of the statistical error in the value of ${\rm
M_{proj}}$. It does not account for any error introduced by incorrect
assumptions about the nature of the cluster orbits.

\citet{tenjes_etal98} quote a total luminosity for M81 of
${\rm (L_{B})_{M81}=1.8\pm0.3\times10^{10}\lsun}$.  Using the mass 
derived above we determine a lower limit for M81's mass-to-light
ratio of $\rm(M/L_B)_{M81}>22$, in solar units. 

\myfig{massfig} shows a rough mass profile for M81 based on masses
derived using clusters inside projected radii of 4, 8, 15 and $22\kpc$.
The specific values are listed in \mytab{masstable}.  The enclosed
mass increases at large radius, indicative of a dark matter halo around M81. 
Evidence for a dark matter halo was also detected by \citet{PBH95}.

\section{Discussion \& Conclusions}
We have obtained spectra of 16 \gc~candidates in M81 with sufficient
signal-to-noise to determine velocities and line-strengths for a
number of absorption features.  We also have supplemented our velocity
data with velocities of 28 additional clusters from previous work, in
order to investigate velocity structures.  Our conclusions are as
follows:
\begin{enumerate}
\item All 16 candidates we observed are confirmed as \bonafide~M81
\gcs.  One appears to have been undergoing a transient event at the
time of observation.
  
\item Our spectroscopic sample is biased to the metal-rich end of the
metallicity range.  Due to the spatial location of the clusters near the
center of the galaxy, the Keck sample is likely to be dominated by bulge
and/or thick-disk clusters.  The former seems more likely, given the
lack of a strong rotation signature in the Keck sample alone, but we
note that rotation could be obscured by spatial bias in such a small
sample.  Based on the spectroscopically-determined metallicities of
clusters in our sample, combined with metallicity estimates from
previous work, we find the metallicity of the entire M81 \GCS~to be
similar to the metallicities of the \MW~and M31 systems.

\item Nearly all the clusters' ages, based on \Hbeta~absorption-line strengths,
are consistent with ages of \MW~and M31 clusters.  However, one cluster in the
sample may be as young as a few Gyrs.  Similar young clusters
have been observed in both M31 and M33.

\item Correlations between absorption line indices which have been
established for \gcs~in the \MW~and M31 hold for the M81 clusters as
well, at least at the upper end of the metallicity-range where the
Keck sample probes.
 
\item Based on the projected mass estimator, the mass of M81 is similar to
the masses of the \MW~and M31.  The projected mass profile of M81 is
consistent with the presence of a dark matter halo.

\item The M81 cluster system appears to be kinematically similar
to the \MW~and M31 systems in that the metal-rich clusters show
evidence for significant rotation while the metal-poor clusters exhibit
halo-like kinematics.  The clusters which are rotating are doing so in
the same sense as M81's disk and the global-rotation signature in our
sample is driven primarily by clusters with \pgcri~between
4 and 8 \kpc.  This may indicate the presence of a thick disk in M81.
A similar analysis of the M31 \GCS~suggests that such a
population is present in that galaxy as well.

It is interesting to note that the annulus of rapidly-rotating,
metal-rich clusters resides at the same galactocentric distance in all
three galaxies discussed here.  One interpretation is that the spatial
and kinematic properties of these globular clusters do not obey galaxy
scaling laws, but then the extent to which various components of the
galaxies themselves follow scaling laws is not yet on firm footing.
The thin (optical) disk of M31 is larger than that of the
\MW~\citep{walterbos_kennicutt88,binney_tremaine_87}, and M31's
metal-rich clusters extend farther than their counterparts in the \MW.
Unfortunately the radial scale length of the thick disk component, with
which the rotating metal-rich clusters are thought to be associated, is
only partially constrained in the Milky Way (see \citet{buser_etal99} and
references therein), and it has not been measured directly for either
M31 or M81.  Also, recent estimates of the dynamical masses of M31 and
the \MW~using satellite galaxies, distant globular clusters and
planetary nebulae as tracers suggest that, in spite of its larger
appearance, M31 may actually be {\it less} massive than the
\MW~\citep{evans_wilkinson_00}.  Obviously more kinematic and
metallicity data are needed for globular clusters in a variety of
spirals to determine if an annulus of rapidly rotating metal-rich
\gcs~are present in all spirals and, if not, whether its presence
depends on any intrinsic or environmental property of the parent galaxy.

\end{enumerate}

In summary, the \GCS~of M81 is both kinematically and chemically very
similar to the systems of the \MW~and M31.  This result lends support to
the idea that universal processes govern the formation of \GCSs~around
spirals.  It may also be relevant to the formation of early-type
galaxies.  As has been discussed extensively \citep{forbes_etal97,AZ98},
it is difficult to explain the \GCSs~of high specific frequency
ellipticals by a simple spiral-merger scenario, but this mechanism could
be relevant in producing low specific frequency ellipticals.  The blue
sub-populations of \gcs~in such galaxies, having originated in the
spiral galaxy progenitors, should be very similar.

\smallskip
\noindent
{\bf Acknowledgments}\\
We thank Karl Gebhardt for the use of his maximum-likelihood code and
many related conversations.  We are grateful to Pauline Barmby for
providing an up-to-date catalogue of velocities and metallicities for
M31 globular clusters.  Also, we thank both Graeme Smith and Francois
Schweizer for carefully reading the paper and making excellent
suggestions.  This work was supported by National Science Foundation
grant number AST990732 and Faculty Research funds of the University of
California, Santa Cruz.

\pagestyle{empty}
\renewcommand{\arraystretch}{0.6}
\begin{deluxetable}{clrrrrcrrr}
\rotate
\tablewidth{0pc}
\tablenum{1}
\tablecaption{Observed Cluster Candidates}
\tablehead{
                  &               &                             &                           & \colhead{$\Delta\alpha$} & \colhead{$\Delta\delta$} & \colhead{$\rm R_{P}$}     &                     &                     &                     \\ 
\colhead{Slitlet} & \colhead{ID}  &  \colhead{$\alpha$(J2000)}  & \colhead{$\delta$(J2000)} & \colhead{(arcmin)}       & \colhead{(arcmin)}       & \colhead{(arcmin)}        & \colhead{{$\rm V$}} & \colhead{$\rm B-V$} & \colhead{$\rm V-R$} \\ 
\colhead{(1)}     & \colhead{(2)} & \colhead{(3)}               & \colhead{(4)}             & \colhead{(5)}            & \colhead{(6)}            & \colhead{(7)}             &\colhead{(8)}        & \colhead{(9)}       & \colhead{(10)}       }
\startdata 
 1  &  50285  & 09:55:55.090   & 69:00:56.14  &  1.908  & $-2.970$  &  3.530 & 18.54  &  1.10  &  0.66  \\
 2  &  50304  & 09:55:49.326   & 69:01:15.29  &  1.391  & $-2.651$  &  2.994 & 18.97  &  1.02  &  0.60  \\
 3  &  50359  & 09:55:37.461   & 69:02:07.58  &  0.329  & $-1.779$  &  1.810 & 18.35  &  1.04  &  0.59  \\
 4  &  50378  & 09:55:57.872   & 69:02:23.07  &  2.154  & $-1.521$  &  2.637 & 19.24  &  1.05  &  0.62  \\
 5  &  50418  & 09:55:54.642   & 69:02:52.57  &  1.865  & $-1.030$  &  2.130 & 18.45  &  1.04  &  0.57  \\
 6  &  50460  & 09:55:51.466   & 69:03:23.60  &  1.580  & $-0.512$  &  1.661 & 18.80  &  0.97  &  0.46  \\
 7  &  50514  & 09:55:48.157   & 69:03:52.03  &  1.284  & $-0.039$  &  1.285 & 19.05  &  1.00  &  0.54  \\
 8  &  50667  & 09:55:22.295   & 69:05:19.16  & $-1.025$  &  1.414  &  1.746 & 18.01  &  1.04  &  0.56  \\
 9  &  50690  & 09:55:21.587   & 69:05:31.98  & $-1.088$  &  1.627  &  1.958 & 18.76  &  0.98  &  0.48  \\
 10  &  50738  & 09:55:30.277   & 69:06:06.17  & $-0.312$  &  2.197  &  2.219 & 20.08  &  0.84  &  0.68  \\
 11  &  50759  & 09:55:35.841   & 69:06:25.53  &  0.184  &  2.520  &  2.526 & 18.59  &  0.82  &  0.40  \\
 12  &  50782  & 09 55 33.042   & 69 06 39.88  & $-0.066$  &  2.759  & 2.760  & 18.70  &  1.00  &  0.52	       \\
 13  &  50787  & 09:56:05.569   & 69:06:42.97  &  2.834  &  2.810  &  3.991 & 19.12  &  0.87  &  0.57  \\
 14  &  50834  & 09:55:25.403   & 69:07:14.72  & $-0.747$  &  3.340  &  3.422 & 19.03  &  0.92  &  0.51  \\
 15  &  50867  & 09:55:51.995   & 69:07:39.32  &  1.622  &  3.749  &  4.085 & 19.83  &  0.75  &  0.30  \\
 16  &  50889  & 09:55:40.194   & 69:07:30.82  &  0.571  &  3.608  &  3.653 & 18.73  &  0.99  &  0.62  \\ 
M81 Center & \nodata & 09:55:33.780 & 69:03:54.34 & 0.000  &  0.000  &  0.000 &\nodata&\nodata&\nodata\\
\enddata
\tablecomments{IDs, coordinates, magnitudes and colors were obtained from Perelmuter \& Racine (1995).}
\tablenotetext{(1)}{Slitlet number on LRIS mask.}
\tablenotetext{(2)}{ID number.}
\tablenotetext{(3,4)}{Equatorial coordinates, J2000.}
\tablenotetext{(5,6)}{Position offsets from M81 center (arcminutes).}
\tablenotetext{(7)}{Projected radius from M81 center (arcminutes).}
\tablenotetext{(8)}{Apparent V magnitude.}
\tablenotetext{(9,10)}{Colors, uncorrected for reddening.}
\label{posntable}
\end{deluxetable}

\pagestyle{empty}
\begin{deluxetable}{clr}
\tablewidth{0pc}
\tablenum{2}
\tablecaption{Velocities for 16 M81 Globular Clusters}
\tablehead{
 &  & \colhead{$v_{los}$} \\
\colhead{Slitlet} & \colhead{ID} & \colhead{($\rm km\,s^{-1}$)} \\
\colhead{(1)} & \colhead{(2)} & \colhead{(3)}
 }
\startdata 
 1  &  50285  & $-199$ (33)  \\
 2  &  50304  &   19   (27)  \\
 3  &  50359  & $-131$   (17)  \\
 4  &  50378  & $-318$   (31)  \\
 5  &  50418  &  $-63$   (33)  \\
 6  &  50460  &   59   (32)  \\
 7  &  50514  & $-335$   (28)  \\
 8  &  50667  & $-144$   (09)  \\
 9  &  50690  & $-261$   (27)  \\
 10  &  50738  &   22   (39)  \\
 11  &  50759  &   11   (24)  \\
 12  &  50782  & $-200$   (17)  \\
 13  &  50787  &  $-87$   (49)  \\
 14  &  50834  &   69   (16)  \\
 15  &  50867  &  $-13$   (29)  \\
 16  &  50889  & $-100$   (25)  
\enddata
\tablenotetext{(1)}{Slitlet number on LRIS mask.}
\tablenotetext{(2)}{ID number from Perelmuter \& Racine (1995).}
\tablenotetext{(3)}{Heliocentric line-of-sight velocity (error in parenthesis).}
\label{veltable}
\end{deluxetable}

\pagestyle{empty}
\begin{deluxetable}{rrrrrrrrrr}
\tablewidth{0pc}\tablenum{3}
\tablecaption{Brodie \& Huchra Metallicity}
\tablehead{ \colhead{Slit}&\colhead{$\rm \Delta$}&\colhead{CNB}&\colhead{CNR}&\colhead{G}&\colhead{MgH}&\colhead{Mg2}&\colhead{Fe52}&\colhead{Na{\sc I}}&\colhead{$\langle\rm{[Fe/H]}\rangle$} \\
\colhead{(1)} & \colhead{(2)} & \colhead{(3)} & \colhead{(4)} & \colhead{(5)} & \colhead{(6)} & \colhead{(7)} & \colhead{(8)} & \colhead{(9)}&\colhead{(10)} }

 \startdata 
1 & \nodata & \nodata & \nodata & \nodata & -1.655  &  -1.194  &  -0.658  &  -0.352  &    -1.207  \\ 
 & \nodata & \nodata & \nodata & \nodata & {\small  0.073} & {\small 0.038} & {\small  0.105} & {\small  0.064} & {\small 0.369} \\
2 & \nodata & \nodata & \nodata & \nodata & -0.460  &  -0.812  &  -0.832  &  -0.776  &    -0.707  \\ 
 & \nodata & \nodata & \nodata & \nodata & {\small  0.102} & {\small 0.054} & {\small  0.148} & {\small  0.094} & {\small 0.167} \\
3 & \nodata & \nodata & \nodata &  0.030  &  -0.513  &  -0.216  &  -0.218  &  -0.407  &    -0.211  \\ 
 & \nodata & \nodata & \nodata & {\small  0.191} & {\small  0.067} & {\small  0.036} & {\small  0.100} & {\small  0.063} & {\small 0.193} \\
4 & \nodata & \nodata & \nodata & \nodata & -0.470  &  -0.311  &  -0.505  &  -0.887  &    -0.407  \\ 
 & \nodata & \nodata & \nodata & \nodata & {\small  0.122} & {\small  0.065} & {\small  0.177} & {\small  0.111} & {\small 0.088} \\
5 & \nodata & \nodata & \nodata & \nodata & -1.210  &  -1.062  &  -0.972  &  -0.821  &    -1.086  \\ 
 & \nodata & \nodata & \nodata & \nodata & {\small  0.072} & {\small  0.038} & {\small  0.108} & {\small  0.067} & {\small 0.091} \\
6 & \nodata & \nodata & \nodata & \nodata & -1.785  &  -1.415  &  -1.262  &  -1.383  &    -1.493  \\ 
 & \nodata & \nodata & \nodata & \nodata & {\small  0.099} & {\small  0.052} & {\small  0.151} & {\small  0.095} & {\small 0.206} \\
7 & \nodata & \nodata & \nodata & \nodata & -1.057  &  -0.971  &  -0.794  &  -1.151  &    -0.955  \\ 
 & \nodata & \nodata & \nodata & \nodata & {\small  0.125} & {\small  0.066} & {\small  0.187} & {\small  0.116} & {\small 0.098} \\
8 &  0.104  &  -0.169  &  -0.706  &  -0.615  &  -0.710  &  -0.744  &  -0.754  &  -1.082  &    -0.698  \\ 
 & {\small  0.057} & {\small  0.350} & {\small  0.070} & {\small  0.122} & {\small  0.053} & {\small  0.028} & {\small  0.079} & {\small  0.047} & {\small 0.058} \\
9 & -0.233  &  -0.717  &  -1.125  &  -1.108  &  -1.254  &  -1.375  &  -1.029  &  -1.488  &    -1.212  \\ 
 & {\small  0.133} & {\small  1.062} & {\small  0.127} & {\small  0.213} & {\small  0.092} & {\small  0.048} & {\small  0.136} & {\small  0.082} & {\small 0.133} \\
10 & \nodata & \nodata & -1.399  &  -1.811  &  -1.618  &  -1.209  &  -0.750  &  -1.286  &    -1.322  \\ 
 & \nodata & \nodata & {\small  0.758} & {\small  0.928} & {\small  0.348} & {\small  0.183} & {\small  0.527} & {\small  0.380} & {\small 0.356} \\
11 & \nodata & \nodata & -2.222  &  -0.558  &  -1.691  &  -1.193  &  -1.193  &  -1.745  &    -1.114  \\ 
 & \nodata & \nodata & {\small  0.092} & {\small  0.175} & {\small  0.069} & {\small  0.036} & {\small  0.104} & {\small  0.070} & 0.409\\
13 & \nodata & \nodata & \nodata & \nodata & -1.074  &  -1.099  &  -0.948  &  -1.568  &    -1.055  \\ 
 & \nodata & \nodata & \nodata & \nodata & {\small  0.149} & {\small  0.078} & {\small  0.208} & {\small  0.135} & {\small 0.062} \\
14 & -0.210  &  -0.315  &  -1.140  &  -1.028  &  -1.179  &  -1.171  &  -1.023  &  -1.385  &    -1.107  \\ 
 & {\small  0.244} & {\small  2.468} & {\small  0.142} & {\small  0.239} & {\small  0.096} & {\small  0.051} & {\small  0.145} & {\small  0.094} & {\small 0.074} \\
15 & \nodata & \nodata & \nodata & -0.080  &  -1.981  &  -1.251  &  -0.400  &  -1.066  &    -1.014  \\ 
 & \nodata & \nodata & \nodata & {\small  0.468} & {\small  0.173} & {\small  0.092} & {\small  0.270} & {\small  0.191} & {\small 0.713} \\
16 & \nodata & \nodata & \nodata & \nodata & -0.687  &  -0.630  &  -0.739  &  -0.804  &    -0.674  \\ 
 & \nodata & \nodata & \nodata & \nodata & {\small  0.100} & {\small  0.053} & {\small  0.145} & {\small  0.091} & {\small 0.044 }
\enddata
\tablecomments{There are two lines for each cluster.  The first line contains the metallicity measurements.  The second line contains the uncertainties in the metallicity measurements.  The uncertainties for the individual metallicity measurements were determined by determining a ``total error'' for the index measurements, (equation 19 of Brodie \& Huchra, 1990) and then propagating that total error through their derived linear index-metallicity relationships.  The uncertainty quoted for the weighted mean metallicity for each cluster is weighted dispersion about weighted mean, where the weights were derived using equation 20 of Brodie \& Huchra, 1990. }
\tablenotetext{(1)}{Slitlet number on LRIS mask.}
\tablenotetext{(2-9)}{Metallicities predicted by various absorption line strengths, as described in Brodie \& Huchra (1990).}
\tablenotetext{(10)}{Error-weighted mean, does not include $\rm\Delta$ or NaI.}
\label{BHmettable}
\end{deluxetable}

\pagestyle{empty}
\begin{deluxetable}{rrrrrrrrr}
\tablewidth{0pc}\tablenum{4}
\tablecaption{Indices Used to Predict Metallicity}
\tablehead{
\colhead{Slitlet} & \colhead{$\Delta$} & \colhead{CNB} & \colhead{CNR} & 
\colhead{G} & \colhead{MgH} & \colhead{Mg2} & \colhead{Fe52} & 
\colhead{Na{\sc I}} \\
\colhead{(1)} & \colhead{(2)} & \colhead{(3)} & \colhead{(4)} &
\colhead{(5)} & \colhead{(6)} & \colhead{(7)} & \colhead{(8)} &
\colhead{(9)}} 
\startdata 
1 & \nodata & \nodata & \nodata & \nodata &  0.009  &   0.103  &   0.070  &     0.139  \\ 
 & \nodata & \nodata & \nodata & \nodata & {\small  0.004} & {\small  0.004} & {\small  0.005} & {\small 0.004} \\
2 & \nodata & \nodata & \nodata & \nodata &  0.067  &   0.141  &   0.062  &     0.110  \\ 
 & \nodata & \nodata & \nodata & \nodata & {\small  0.005} & {\small  0.005} & {\small  0.007} & {\small 0.007} \\
3 & \nodata & \nodata & \nodata &  0.218  &   0.064  &   0.201  &   0.092  &     0.135  \\ 
 & \nodata & \nodata & \nodata & {\small  0.017} & {\small  0.003} & {\small  0.004} & {\small  0.005} & {\small 0.004} \\
4 & \nodata & \nodata & \nodata & \nodata &  0.067  &   0.192  &   0.078  &     0.102  \\ 
 & \nodata & \nodata & \nodata & \nodata & {\small  0.006} & {\small  0.007} & {\small  0.009} & {\small 0.008} \\
5 & \nodata & \nodata & \nodata & \nodata &  0.031  &   0.116  &   0.055  &     0.106  \\ 
 & \nodata & \nodata & \nodata & \nodata & {\small  0.004} & {\small  0.004} & {\small  0.005} & {\small 0.005} \\
6 & \nodata & \nodata & \nodata & \nodata &  0.003  &   0.080  &   0.040  &     0.068  \\ 
 & \nodata & \nodata & \nodata & \nodata & {\small  0.005} & {\small  0.005} & {\small  0.007} & {\small 0.007} \\
7 & \nodata & \nodata & \nodata & \nodata &  0.038  &   0.125  &   0.063  &     0.084  \\ 
 & \nodata & \nodata & \nodata & \nodata & {\small  0.006} & {\small  0.007} & {\small  0.009} & {\small 0.008} \\
8 &  0.796  &   0.269  &   0.066  &   0.161  &   0.055  &   0.148  &   0.065  &     0.088  \\ 
 & {\small  0.018} & {\small  0.053} & {\small  0.010} & {\small  0.011} & {\small  0.003} & {\small  0.003} & {\small  0.004} & {\small 0.003} \\
9 &  0.690  &   0.186  &   0.009  &   0.118  &   0.028  &   0.084  &   0.052  &     0.060  \\ 
 & {\small  0.042} & {\small  0.162} & {\small  0.017} & {\small  0.019} & {\small  0.004} & {\small  0.005} & {\small  0.007} & {\small 0.006} \\
10 & \nodata & \nodata & -0.028  &   0.056  &   0.011  &   0.101  &   0.066  &     0.074  \\ 
 & \nodata & \nodata & {\small  0.103} & {\small  0.081} & {\small  0.017} & {\small  0.018} & {\small  0.026} & {\small 0.026} \\
11 & \nodata & \nodata & -0.140  &   0.166  &   0.007  &   0.103  &   0.044  &     0.043  \\ 
 & \nodata & \nodata & {\small  0.013} & {\small  0.015} & {\small  0.003} & {\small  0.004} & {\small  0.005} & {\small 0.005} \\
13 & \nodata & \nodata & \nodata & \nodata &  0.037  &   0.112  &   0.056  &     0.055  \\ 
 & \nodata & \nodata & \nodata & \nodata & {\small  0.007} & {\small  0.008} & {\small  0.010} & {\small 0.009} \\
14 &  0.698  &   0.247  &   0.007  &   0.125  &   0.032  &   0.105  &   0.052  &     0.067  \\ 
 & {\small  0.077} & {\small  0.375} & {\small  0.019} & {\small  0.021} & {\small  0.005} & {\small  0.005} & {\small  0.007} & {\small 0.006} \\
t15 & \nodata & \nodata & \nodata &  0.208  &  -0.007  &   0.097  &   0.083  &     0.090  \\ 
 & \nodata & \nodata & \nodata & {\small  0.041} & {\small  0.008} & {\small  0.009} & {\small  0.013} & {\small 0.013} \\
t16 & \nodata & \nodata & \nodata & \nodata &  0.056  &   0.159  &   0.066  &     0.108  \\ 
 & \nodata & \nodata & \nodata & \nodata & {\small  0.005} & {\small  0.005} & {\small  0.007} & {\small 0.006 }
\enddata
\tablecomments{There are two lines for each cluster.  The first line contains the index measurements.  The second line contains the uncertainty (photon error) in the index measurements. Col.(1): Slitlet number on LRIS mask.  Cols.(2)-(9): Absorption-line index values, measured as prescribed in Brodie \& Huchra 1990.  Uncertainties measured as described in text. }
\label{BHindtable}
\end{deluxetable}

\pagestyle{empty}
\begin{deluxetable}{lllrrrrrrrr}
\rotate
\tablewidth{0pc}
\tablenum{5}
\tablecaption{Data for Additional M81 Clusters}
\tablehead{ 
 & & & \colhead{$\Delta\alpha$} & \colhead{$\Delta\delta$} & \colhead{$R_{P}$} & & & & \colhead{$v_{los}$} & \\ \colhead{ID}  & \colhead{$\alpha$(J2000)} & \colhead{$\delta$(J2000)} & \colhead{(arcmin)}  & \colhead{(arcmin)}  & \colhead{(arcmin)} & \colhead{${\rm V}$} & \colhead{${\rm B-V}$}& \colhead{${\rm V-R}$}&\colhead{$\rm (km\,s^{-1})$}& \colhead{[Fe/H]} \\
\colhead{(1)} & \colhead{(2)} & \colhead{(3)} & \colhead{(4)} &\colhead{(5)} & \colhead{(6)} & \colhead{(7)} & \colhead{(8)} & \colhead{(9)} &\colhead{(10)} &\colhead{(11)} }
\startdata 
   R9 = HS18 = $50785^{**}$ &  9:54:35.88 & 69:06:43.1 & $-5.161$ &   2.812 &  5.878 & 19.08 &  0.86 & 0.56 & 229 (66) & $-1.00$ (0.84) \\
     HS1 = 40312     &  9:54:52.01 & 69:19:45.3 & $-3.686$ &  15.849 & 16.272 & 19.11 &  0.81 & 0.55 & 258 (27) & $-2.10$ (0.97) \\
     HS6 = 60072     &  9:55:08.08 & 68:49:56.5 & $-2.320$ & $-13.964$ & 14.156 & 19.61 &  0.80 & 0.48 & 122 (77) & $-1.77$ (0.83) \\
    HS2 = $50864^*$     &  9:56:25.74 & 69:07:27.8 &  4.629 &   3.558 &  5.838 & 19.62 &  0.26 & 0.26 & 180 (94) & $-2.09$ (0.59) \\
    R12 = $60126^*$     &  9:55:26.97 & 68:44:53.6 & $-0.617$ & $-19.013$ & 19.023 & 19.16 &  0.99 & 0.59 & $-37$ (39) &  0.19 (1.44) \\
     R14 = HS26 = 60012&  9:55:42.09 & 68:55:00.6 &  0.747 &  $-8.896$ &  8.927 & 20.01 &  0.97 & 0.62 &$-129$ (57) &  0.40 (2.65) \\
   HS35 = $50960$    &  9:55:52.06 & 69:08:18.7 &  1.627 &   4.406 &  4.697 & 18.49 &  0.86 & 0.54 & 180 (59) & $-1.43$ (0.40) \\
  HS19 = $50415^{**}$    &  9:56:20.58 & 69:02:49.0 &  4.184 &  $-1.089$ &  4.323 & 19.24 &  0.85 & 0.48 &$-218$ (51) & $-1.46$ (0.52) \\
\enddata
\tablenotetext{*}{Likely to be a Galactic star, rejected from the sample.}
\tablenotetext{**}{Also observed by Perelmuter et. al, 1995 (See \linesec.}
\tablenotetext{(1)}{ID from Perelmuter \& Racine (1995).}
\tablenotetext{(2,3)}{Equatorial Coordinates (J2000).}
\tablenotetext{(4,5)}{Position offsets from M81 center (arcminutes)}
\tablenotetext{(6)}{Projected radius from M81 center (arcminutes).}
\tablenotetext{(7)}{Apparent V magnitude.}
\tablenotetext{(8,9)}{Colors, uncorrected for reddening.}
\tablenotetext{(10)}{Heliocentric, line-of-sight velocity (J. Huchra, private communication). Errors in parentheses.}
\tablenotetext{(11)}{Metallicity from Brodie \& Huchra (1991).}
\label{johntable}
\end{deluxetable}

\pagestyle{empty}
\begin{deluxetable}{rrrrrrrrrrrrr}
\rotate
\tablewidth{0pc}\tablenum{6a}
\tablecaption{Lick IDS Index Values}
\tablehead{
\colhead{Slit}&\colhead{CN1}&\colhead{CN2}&\colhead{Ca4227}&\colhead{G4300}&\colhead{Fe4383}&\colhead{Ca4455}&\colhead{Fe4531}&\colhead{Fe4668}&\colhead{H$\beta$}&\colhead{Fe5015}&\colhead{Mg1}&\colhead{Mg2}\\
\colhead{(1)}&\colhead{(2)}&\colhead{(3)}&\colhead{(4)}&\colhead{(5)}&\colhead{(6)}&\colhead{(7)}&\colhead{(8)}&\colhead{(9)}&\colhead{(10)}&\colhead{(11)}&\colhead{(12)}&\colhead{(13)}
}
\startdata 
1 & \nodata & \nodata & \nodata & \nodata & \nodata & \nodata & \nodata & \nodata &  1.967  &  3.500  &   0.011 &  0.10  \\
  & \nodata & \nodata & \nodata & \nodata & \nodata & \nodata & \nodata & \nodata & {0.197} & {0.414} & {0.004} & {0.004} \\
2 & \nodata & \nodata & \nodata & \nodata & \nodata & \nodata & \nodata &  0.790  &   1.903 &  3.688  &  0.068  &  0.139 \\
 & \nodata & \nodata & \nodata & \nodata & \nodata & \nodata & \nodata & {0.950} & {0.268} & {0.578} & {0.005} & {0.005} \\
3 & \nodata & \nodata & \nodata &  6.651  &   3.510  &   2.012  &   6.974  &   4.097  &   1.929  &   5.226  &   0.066  &   0.199 \\
 & \nodata & \nodata & \nodata & {0.419} & {0.600} & {0.267} & {0.374} & {0.576} & {0.175} & {0.372} & {0.003} & {0.004} \\
4 & \nodata & \nodata & \nodata & \nodata & \nodata & \nodata & \nodata & \nodata &  2.246  &   5.447  &   0.069  &   0.188 \\ 
 & \nodata & \nodata & \nodata & \nodata & \nodata & \nodata & \nodata & \nodata & {0.318} & {0.669} & {0.006} & {0.006} \\
5 & \nodata & \nodata & \nodata & \nodata & \nodata & \nodata & \nodata & \nodata &  1.868  &   3.357  &   0.031  &   0.114 \\ 
 & \nodata & \nodata & \nodata & \nodata & \nodata & \nodata & \nodata & \nodata & {0.187} & {0.407} & {0.003} & {0.004} \\
6 & \nodata & \nodata & \nodata & \nodata & \nodata & \nodata & \nodata &  3.083  &   2.278  &   2.166  &   0.002  &   0.080 \\
 & \nodata & \nodata & \nodata & \nodata & \nodata & \nodata & \nodata & {0.852} & {0.249} & {0.563} & {0.005} & {0.005} \\
7 & \nodata & \nodata & \nodata & \nodata & \nodata & \nodata &  4.119  &   3.279  &   1.832  &   2.684  &   0.040  &   0.123 \\
 & \nodata & \nodata & \nodata & \nodata & \nodata & \nodata & {0.700} & {1.058} & {0.324} & {0.726} & {0.006} & {0.007} \\
8 &  0.068  &   0.105  &   1.090  &   4.955  &   3.919  &   1.281  &   2.942  &   2.128  &   1.898  &   3.986  &   0.055  &   0.146 \\
 & {0.009} & {0.012} & {0.171} & {0.284} & {0.411} & {0.188} & {0.294} & {0.446} & {0.135} & {0.292} & {0.003} & {0.003} \\
9 &  0.013  &   0.057  &   0.256  &   3.675  &   1.721  &   1.102  &   2.465  &   1.040  &   1.807  &   3.215  &   0.029  &   0.084 \\
 & {0.017} & {0.023} & {0.311} & {0.520} & {0.773} & {0.337} & {0.526} & {0.810} & {0.237} & {0.514} & {0.004} & {0.005} \\
10 & -0.038  &  -0.015  &   0.668  &   1.484  &   4.887  &   1.275  &   1.956  &   2.117  &   1.769  &   6.273  &   0.010  &   0.100 \\
 & {0.102} & {0.127} & {1.623} & {2.420} & {2.811} & {1.262} & {1.973} & {2.949} & {0.896} & {1.842} & {0.017} & {0.018} \\
11 & -0.135  &  -0.103  &   1.180  &   5.530  &   4.098  &   1.604  &   2.696  &   2.308  &   1.766  &   2.691  &   0.007  &   0.097 \\
 & {0.012} & {0.016} & {0.211} & {0.404} & {0.539} & {0.237} & {0.391} & {0.579} & {0.177} & {0.394} & {0.003} & {0.004} \\
13 & \nodata & \nodata & \nodata & \nodata & \nodata & \nodata & \nodata & \nodata & \nodata &  2.946  &   0.036  &   0.112 \\
 & \nodata & \nodata & \nodata & \nodata & \nodata & \nodata & \nodata & \nodata & \nodata & {0.850} & {0.007} & {0.008} \\
14 &  0.004  &   0.035  &   0.544  &   3.773  &   1.394  &   0.834  &   2.657  &   1.540  &   2.167  &   3.116  &   0.032  &   0.104 \\
 & {0.019} & {0.026} & {0.343} & {0.582} & {0.850} & {0.363} & {0.552} & {0.845} & {0.250} & {0.547} & {0.005} & {0.005} \\
15 & \nodata & \nodata &  1.178  &   6.072  &   7.390  &   1.291  &   4.693  &  -0.874  &   4.509  &   6.897  &  -0.005  &   0.099 \\
 & \nodata & \nodata & {0.649} & {1.050} & {1.457} & {0.711} & {1.051} & {1.544} & {0.407} & {0.888} & {0.008} & {0.009} \\
16 & \nodata & \nodata & \nodata & \nodata & \nodata & \nodata &  3.681  &   5.688  &   1.643  &   4.092  &   0.057  &   0.157 \\
 & \nodata & \nodata & \nodata & \nodata & \nodata & \nodata & {0.689} & {0.891} & {0.270} & {0.560} & {0.005} & {0.005} 
\enddata
\tablenotetext{(1)}{Slitlet number on LRIS mask.}
\tablenotetext{(2-13)}{Absorbption-line index values measured as described in Trager 1998 and references therein.}
\label{LIDStablea}
\end{deluxetable}

\pagestyle{empty}
\begin{deluxetable}{rrrrrrrrrrrrrr}
\rotate
\tablewidth{0pc}\tablenum{6b}
\tablecaption{Lick IDS Index Values, continued}
\tablehead{
Slit&\colhead{Mgb}&\colhead{Fe5270}&\colhead{Fe5335}&\colhead{Fe5406}&\colhead{Fe5709}&\colhead{Fe5782}&\colhead{NaD}&\colhead{TiO1}&\colhead{TiO2}&\colhead{H$\delta$F}&\colhead{H$\gamma$F}&\colhead{H$\delta$A}&\colhead{H$\delta$A}
\\
\colhead{(1)}&\colhead{(14)}&\colhead{(15)}&\colhead{(16)}&\colhead{(17)}&\colhead{(18)}&\colhead{(19)}&\colhead{(20)}&\colhead{(21)}&\colhead{(22)}&\colhead{(23)}&\colhead{(24)}&\colhead{(25)}&\colhead{(26)}
}
\startdata 
1 &   2.936  &   2.529  &   2.063  &   1.269  &   0.785  &   0.762  &   3.765  &   0.021  &   0.056  &  \nodata & \nodata & \nodata &\nodata \\
  & {0.167} & {0.173} & {0.217} & {0.149} & {0.097} & {0.094} & {0.113} & {0.003} & {0.002} & \nodata & \nodata & \nodata & \nodata \\
2 &   2.695  &   2.159  &   2.104  &   1.026  &   0.783  &   0.547  &   3.026  &   0.032  &   0.027  &  \nodata & \nodata & \nodata &\nodata \\
  & {0.239} & {0.246} & {0.305} & {0.213} & {0.140} & {0.137} & {0.171} & {0.004} & {0.003} & \nodata & \nodata & \nodata & \nodata \\
3 & 3.927  &   3.351  &   3.059  &   1.633  &   0.941  &   0.822  &   3.604  &   0.037  &   0.078  &  \nodata & -1.112  &  \nodata &   -4.977  \\ 
 & {0.156} & {0.161} & {0.201} & {0.142} & {0.095} & {0.092} & {0.112} & {0.003} & {0.002} & \nodata & {0.255} & \nodata & 0.462\\
4 & 4.224  &   2.901  &   2.077  &   1.675  &   0.841  &   0.969  &   2.893  &   0.025  &   0.060  &  \nodata & \nodata & \nodata &\nodata \\
 & {0.274} & {0.289} & {0.356} & {0.245} & {0.167} & {0.160} & {0.203} & {0.005} & {0.004} & \nodata & \nodata & \nodata & \nodata \\
5 &   2.444  &   1.903  &   1.858  &   0.754  &   0.640  &   0.682  &   2.933  &   0.019  &   0.028  &  \nodata & \nodata & \nodata &\nodata \\
 & {0.171} & {0.180} & {0.226} & {0.159} & {0.104} & {0.100} & {0.122} & {0.003} & {0.002} & \nodata & \nodata & \nodata & \nodata \\
6 &   1.687  &   1.398  &   1.507  &   0.855  &   0.351  &   0.267  &   1.890  &   0.012  &   0.004  &  \nodata & \nodata & \nodata &\nodata \\
 & {0.238} & {0.255} & {0.323} & {0.224} & {0.149} & {0.145} & {0.180} & {0.004} & {0.003} & \nodata & \nodata & \nodata & \nodata \\
7 &   2.737  &   2.233  &   2.435  &   1.441  &   0.770  &   0.513  &   2.326  &   0.037  &   0.046  &  \nodata & \nodata & \nodata &\nodata \\
 & {0.293} & {0.312} & {0.387} & {0.268} & {0.178} & {0.174} & {0.215} & {0.005} & {0.004} & \nodata & \nodata & \nodata & \nodata \\
8 &   2.635  &   2.387  &   2.081  &   1.133  &   0.780  &   0.585  &   2.442  &   0.030  &  \nodata &  1.090  &  -0.342  &  -0.377  &    -2.906  \\ 
 & {0.125} & {0.130} & {0.162} & {0.113} & {0.073} & {0.071} & {0.087} & {0.002} & \nodata & {0.242} & {0.164} & {0.381} & 0.288\\
9 &   1.644  &   1.812  &   1.339  &   0.777  &   0.529  &   0.381  &   1.689  &   0.012  &  \nodata &  1.854  &   1.177  &   0.730  &    -0.410  \\ 
 & {0.219} & {0.228} & {0.294} & {0.205} & {0.128} & {0.127} & {0.156} & {0.004} & \nodata & {0.437} & {0.273} & {0.685} & 0.479\\
10 &   1.858  &   1.848  &   0.942  &   0.660  &   0.481  &   0.693  &   2.012  &   0.046  &   0.025  &   0.274  &   1.891  &  -0.310  &     0.700  \\ 
 & {0.840} & {0.883} & {1.174} & {0.804} & {0.530} & {0.508} & {0.712} & {0.015} & {0.013} & {2.700} & {1.155} & {5.514} & 2.054\\
11 &   3.463  &   1.720  &   0.905  &   0.396  &   0.556  &   0.063  &   1.103  &   0.011  &  -0.014  &  \nodata & -0.896  &  \nodata &   -3.318  \\ 
 & {0.156} & {0.174} & {0.226} & {0.158} & {0.103} & {0.105} & {0.135} & {0.003} & {0.003} & \nodata & {0.232} & \nodata & 0.409\\
13 &   2.486  &   2.031  &   1.761  &   1.113  &   0.558  &   0.300  &   1.528  &   0.015  &   0.027  &  \nodata & \nodata & \nodata &\nodata \\
 & {0.348} & {0.346} & {0.432} & {0.302} & {0.197} & {0.195} & {0.257} & {0.005} & {0.005} & \nodata & \nodata & \nodata & \nodata \\
14 &   1.498  &   1.792  &   1.534  &   0.920  &   0.711  &   0.401  &   1.715  &  -0.005  &  \nodata &  1.995  &   0.994  &   1.395  &     0.702  \\ 
 & {0.236} & {0.242} & {0.306} & {0.212} & {0.141} & {0.140} & {0.177} & {0.004} & \nodata & {0.483} & {0.307} & {0.749} & 0.525\\
15 &   0.696  &   2.455  &   2.548  &   0.476  &  -0.032  &   0.257  &   2.264  &   0.021  &  -0.041  &  \nodata &  1.400  &  \nodata &   -5.866  \\ 
 & {0.448} & {0.449} & {0.554} & {0.394} & {0.264} & {0.258} & {0.356} & {0.008} & {0.007} & \nodata & {0.626} & \nodata & 1.273\\
16 &   3.358  &   2.321  &   2.077  &   1.569  &   0.783  &   0.833  &   3.027  &   0.033  &   0.055  &  \nodata & \nodata & \nodata &\nodata \\
 & {0.233} & {0.240} & {0.301} & {0.205} & {0.133} & {0.127} & {0.166} & {0.004} & {0.003} & \nodata & \nodata & \nodata & \nodata 
\enddata
\tablenotetext{(1)}{Slitlet number on LRIS mask.}
\tablenotetext{(14-26)}{Absorbption-line index values measured as described in Trager 1998 and references therein.}
\label{LIDStableb}
\end{deluxetable} 

\pagestyle{empty}
\begin{deluxetable}{rcrcrrrr}
\rotate
\tablewidth{0pc}
\tablenum{7}
\tablecaption{Kinematics of M81 Globular Clusters}
\tablehead{
  & 	&   \colhead{$\langle \rm R_{P}\rangle$} & \colhead{Color}&
     \colhead{$\langle v_{los}\rangle $} & 
     \colhead{$\sigma_{v_{los}}$} & \colhead{$v_{rot}$} &
     \colhead{${\rm \theta_{rot}}$} \\
 \colhead{N} & \colhead{${\rm R_{P} }$ Range} & \colhead{(kpc)} &
     \colhead{Range}  & \colhead{(${\rm km\,s^{-1}}$)} &
     \colhead{(${\rm km\,s^{-1}}$)} & \colhead{(${\rm km\,s^{-1}}$)} &
     \colhead{($^{\rm\circ}$)}  \\
 \colhead{(1)} & \colhead{(2)} &
     \colhead{(3)}  & \colhead{(4)} &
     \colhead{(5)}  & \colhead{(6)} &
     \colhead{(7)}  & \colhead{(8)} }
\startdata		
44&Full Range                &  7.9 &  All               & $-34$ (23) & 133 (19) &  85 (32) &   17 (28) \\
33&Full Range	             &  5.8 & ${\rm B-R}>1.3$    & $-48$ (28) & 143 (25) &  93 (38) &   18 (33) \\
11&Full Range		     & 12.4 & ${\rm B-R}<1.3$    & $-11$ (15) &  11 (48) &  21 (26) &  143 (52) \\
18&Full Range		     &  5.2 & ${\rm B-R}>1.5$    &$-104$ (31) & 121 (37) & 128 (62) &$-11$ (39) \\
16&${\rm R_{pgc}<4\,kpc}$    &  2.8 &  All               &$-103$ (31) & 135 (42) &  89 (84) &$-17$ (35) \\
22&${\rm 2<R_{pgc}<6\,kpc}$  &  3.6 &  All               & $-56$ (33) & 120 (25) & 119 (52) & 14 (28) \\
17&${\rm 3<R_{pgc}<7\,kpc}$  &  4.2 &  All               & $-11$ (38) & 103 (30) & 157 (37) & 27 (20) \\
13&${\rm 4<R_{pgc}<8\,kpc}$  &  5.3 &  All               &   1 (46) & 109 (37) & 168 (42) & 45 (25) \\
15&${\rm R_{pgc}>8\,kpc}$    & 14.2 &  All               &  13 (34) & 125 (30) &  49 (52) & 31 (88) \\
\enddata
\tablenotetext{(1)}{Number of clusters in sample.}
\tablenotetext{(2)}{Spatial selection criteria for sample.}
\tablenotetext{(3)}{Average ${\rm R_{P}}$ of sample (assumes distance to M81 of 3.7 Mpc).}
\tablenotetext{(4)}{Color selection criteria for sample.}
\tablenotetext{(5)}{Mean heliocentric line-of-sight velocity of sample (errors in parentheses).}
\tablenotetext{(6)}{Velocity dispersion, corrected for rotation (errors in parentheses).}
\tablenotetext{(7)}{Amplitude of sample rotation (errors in parenthesis).}
\tablenotetext{(8)}{Best fit position angle of rotation axis, positive angle is east of north. Errors in parentheses. }
\label{kintable}
\end{deluxetable}

\pagestyle{empty}
\begin{deluxetable}{rrrrr}
\tablewidth{0pc}
\tablenum{8}
\tablecaption{Kinematics of M31 Globular Clusters}
\tablehead{  \colhead{Spatial} &  \colhead{$v_{rot}$} &
               \colhead{$\sigma_{v_{los}}$}&  \colhead{${\rm \theta_{rot}}$} &\\
 \colhead{Group} & \colhead{(${\rm km\,s^{-1}}$)} &
     \colhead{(${\rm km\,s^{-1}}$)} & \colhead{($^{\rm\circ}$)}  & N\\
 \colhead{(1)} & \colhead{(2)} &
     \colhead{(3)}  & \colhead{(4)} &
     \colhead{(5)}  }
\startdata
${\bf\rm R_{pgc}<2\,kpc}$ & & & & \\
MR & 96 (56) & 152 (36)  & 109 (56) & 17 \\
MP & 81 (60) & 153 (68)  &$-113$ (51) & 15 \\
$\bf\rm 2<R_{pgc}<6\,kpc$  & &  & \\
MR & 127 (40) & 122 (40) & 119  (24) & 16 \\
MP & 20  (28) & 134 (15) & 164 (108)& 45 \\
$\bf\rm 3<R_P<7\,kpc$ & & & & \\
MR & 143 (48) & 119 (43) & 128 (22) & 14 \\
MP & 37  (33) & 144 (16) & 136 (54) & 43 \\
$\bf\rm 4<R_P<8\,kpc$ & & & & \\
MR & 144 (68) & 129 (50) &  87 (39) & 11 \\
MP &  65 (34) & 139 (21) & 121 (31) & 39 \\
$\bf\rm R_P<8\,kpc$ & & & & \\
MR &  49 (39) & 56  (27) & 96 (44) & 12 \\
MP &  72 (26) & 114 (13) & 98 (29) & 36 \\
\enddata
\tablenotetext{(1)}{Limits of annulus defining sample.  Assumes distance of
744 kpc to M31.}
\tablenotetext{(2)}{Best fit rotation velocity, errors in parentheses}
\tablenotetext{(3)}{Velocity dispersion (corrected for rotation), errors in parentheses.}
\tablenotetext{(4)}{Best fit rotation axis (positive angle is east of north), errors in parenthesis}
\tablenotetext{(5)}{Number of clusters in sample.}
\tablecomments{MR=Metal-rich clusters, those with $\rm [Fe/H]>=-0.8$.  MP indicates clusters with metallicity less than this value.}
\label{m31table}
\end{deluxetable}

\pagestyle{empty}
\begin{deluxetable}{ r r c}
\tablewidth{0pc}
\tablenum{9}
\tablecaption{Projected Mass of M81}
\tablehead{              & \colhead{${\rm R_{max}}$} & \colhead{${\rm M_{proj}}$} \\ 
              \colhead{N} & \colhead{(kpc)}     &\colhead{($10^{11}\msun$)} \\
            \colhead{(1)}       & \colhead{(2)}& \colhead{(3)} }
\startdata
15 &  3.9 & 1.2 (0.8) \\
30 &  7.8 & 2.4 (0.7) \\
38 & 14.7 & 3.1 (0.7) \\
46 & 21.3 & 4.0 (0.8) \\
\enddata
\tablenotetext{(1)}{Number of points in sample.}
\tablenotetext{(2)}{Maximum projected galactocentric radius in sample.}
\tablenotetext{(3)}{Projected mass interior to ${\rm R_{max}}$.}
\tablenotetext{(4)}{Statistical uncertainty in projected mass.  Error
(in parentheses) does not include errors potentially introduced by
assumptions about the cluster orbits.}
\label{masstable}
\end{deluxetable}
\begin{onecolumn}
\eject
\begin{figure}
\plotone{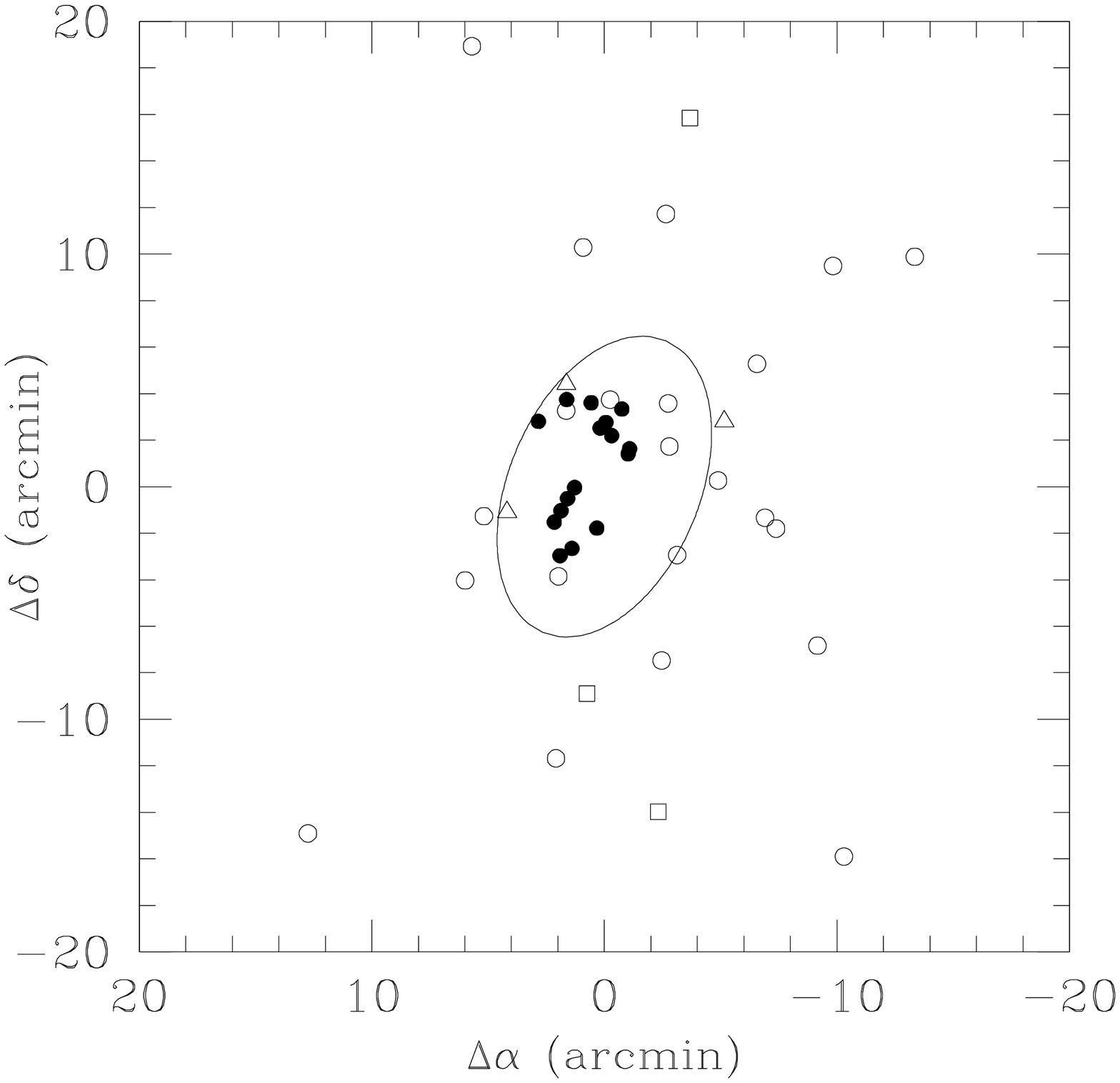}
\figcaption[posfig.eps]{\label{posfig} Positions of all
spectroscopically-confirmed M81 \gcs, relative to the center of M81.
The ellipse shows the location and orientation of M81's disk, taken from
\citet{PBH95}  Solid circles are members of the Keck sample.  Open
circles are clusters studied by \citet{PBH95}, and open squares are
clusters from \citet{BH91} for which velocities have been measured (J.
Huchra, private communication.)  Open triangles indicate clusters 
studied by both \citet{BH91} and \citet{PBH95}.}
\end{figure}

\eject
\begin{figure}
\plotone{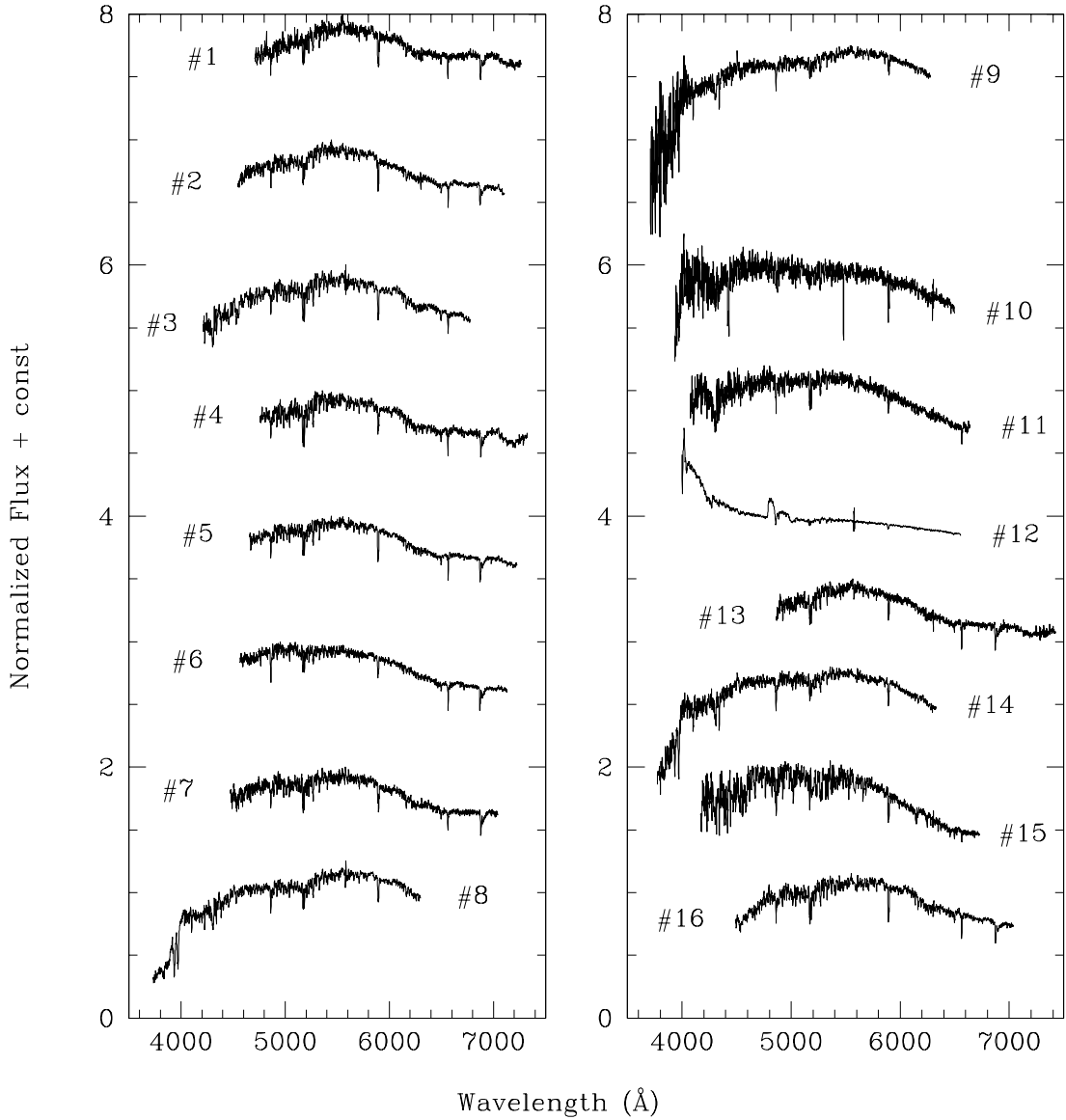}
\figcaption[relspec.eps]{\label{specfig} Spectra of 16 M81
\gc~candidates, obtained with 4500 seconds of integration with LRIS
on the Keck I telescope.  The spectra have been corrected for
foreground reddening and shifted to the rest frame.  Continuum
shapes are reliable only between 3900\AA~and 6500\AA.  All 
are confirmed as M81 \gcs, although object 12 appears to have been observed
during a transient event in the cluster (see \datasec).}
\end{figure}

\eject
\begin{figure}
\plotone{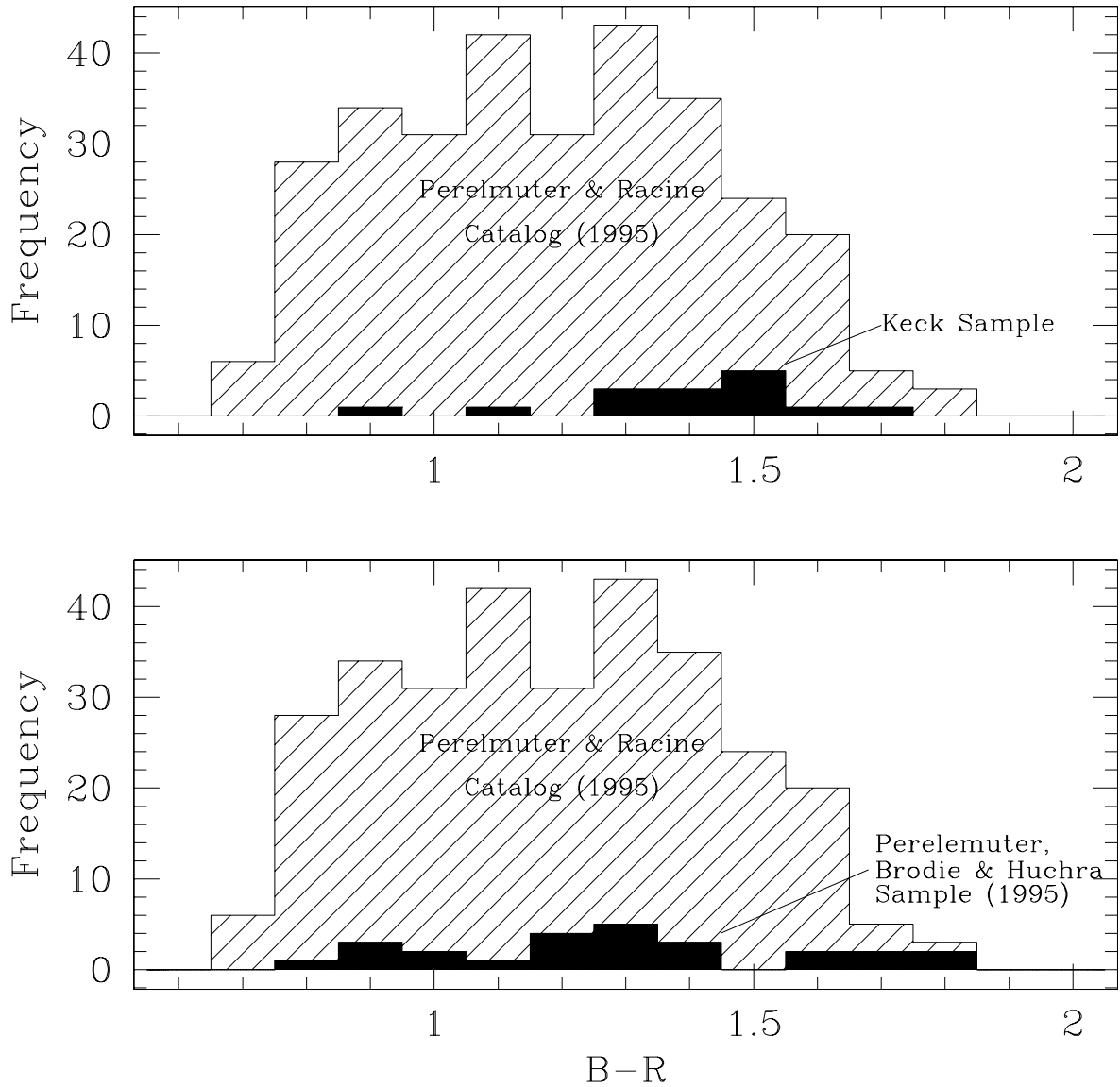}
\figcaption[colorhist.eps]{\label{colorhistfig}
The \color{B}{R} color distribution of the Keck sample (upper panel) and
the sample of \citet{PBH95} (lower panel) compared to the color
distribution of the entire list of M81 cluster candidates from
\citet{PR95}.  The Keck sample is clearly biased to the redder,
metal-rich end of the spectrum.  The \citet{PBH95} sample has a more
uniform color distribution (see \redsec).}
\end{figure}

\eject
\begin{figure}
\plotone{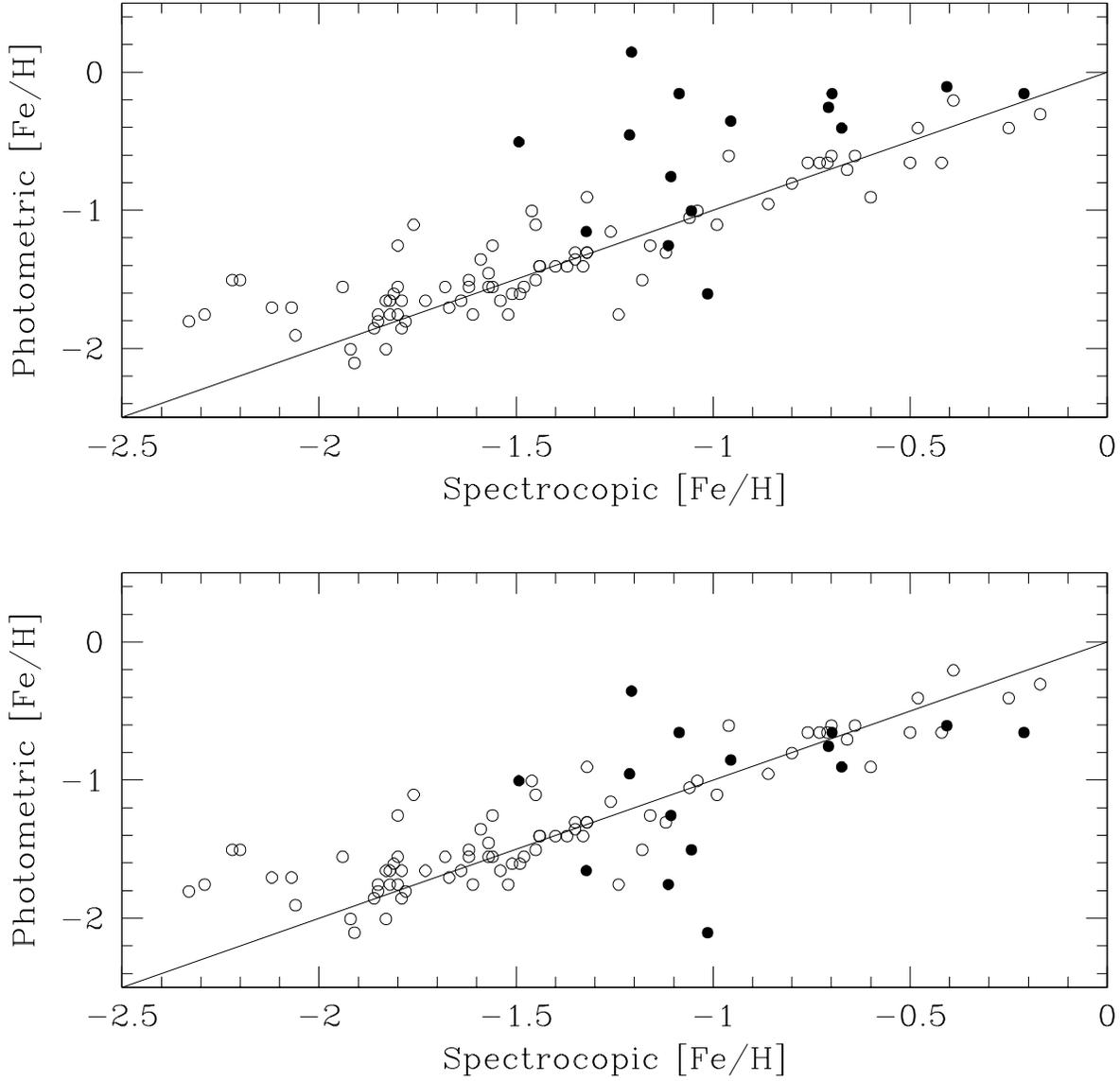}
\figcaption[comparemet.eps]{\label{comparemetfig} Photometric vs.
spectroscopic metallicities for M81 (filled circles) and Milky Way (open
circles) globular clusters.  Photometric metallicities were computed
from \color{B}{V} colors using the color-metallicity derived in
\citet{couture90}.  In the upper panel, the colors were corrected for
foreground reddening only, which was estimated to be $\sim\!0.1$
\citep{PR95}.  The lower panel shows that de-reddening the M81 cluster
colors by an additional 0.1 mag brings the photometric metallicities of
the M81 clusters in line with those of the Milky Way (but see \redsec).
Only Milky Way clusters for which foreground reddening is low
(\excess{B}{V}$<0.4$) are included.  The reddening and metallicity
values for the \MW~clusters were obtained from \citet{harris96}. 
}
\end{figure}

\eject
\begin{figure}
\plotone{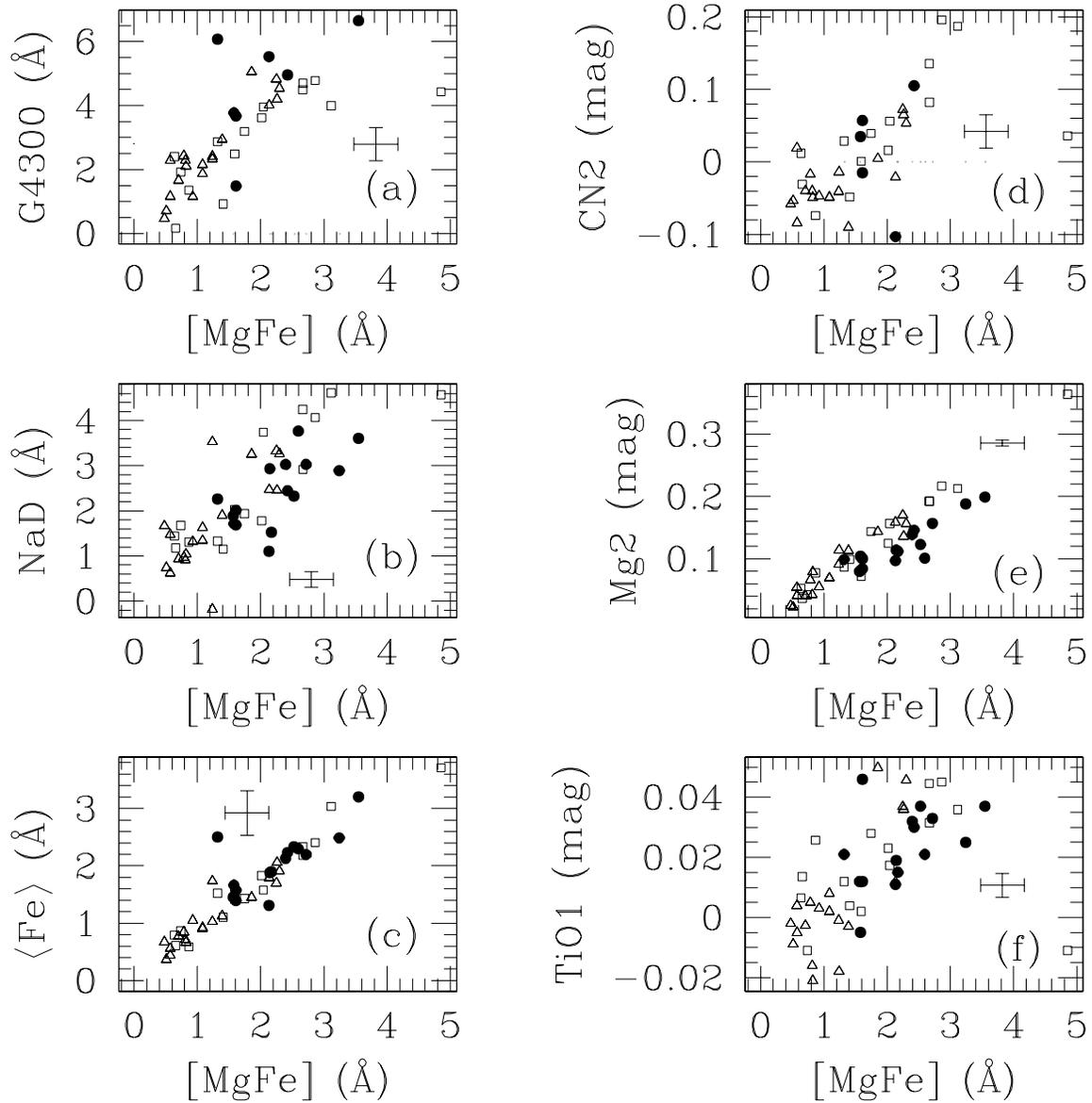}
\figcaption[element.eps]{ \label{elementfig} Selected Lick IDS
indices plotted against the metallicity-sensitive index [MgFe] for
M81 (filled circles), M31 (open squares) and \MW~(open triangles)
\gcs.  The error bars show the typical sizes of the statistical (photon)
errors on the M81 cluster indices.  The correlations between specific
indices observed for Milky Way and M31 globular clusters are also
present in the M81 globular cluster system.  The apparent depression of
Mg2 with respect to [MgFe] (panel {\it e}) is likely due to
contamination of the Fe5335 index.  See \linesec~and \myfig{ironfig} for
more detail.}
\end{figure}

\eject
\begin{figure}
\epsfxsize=14.0cm
\centerline{\epsfbox{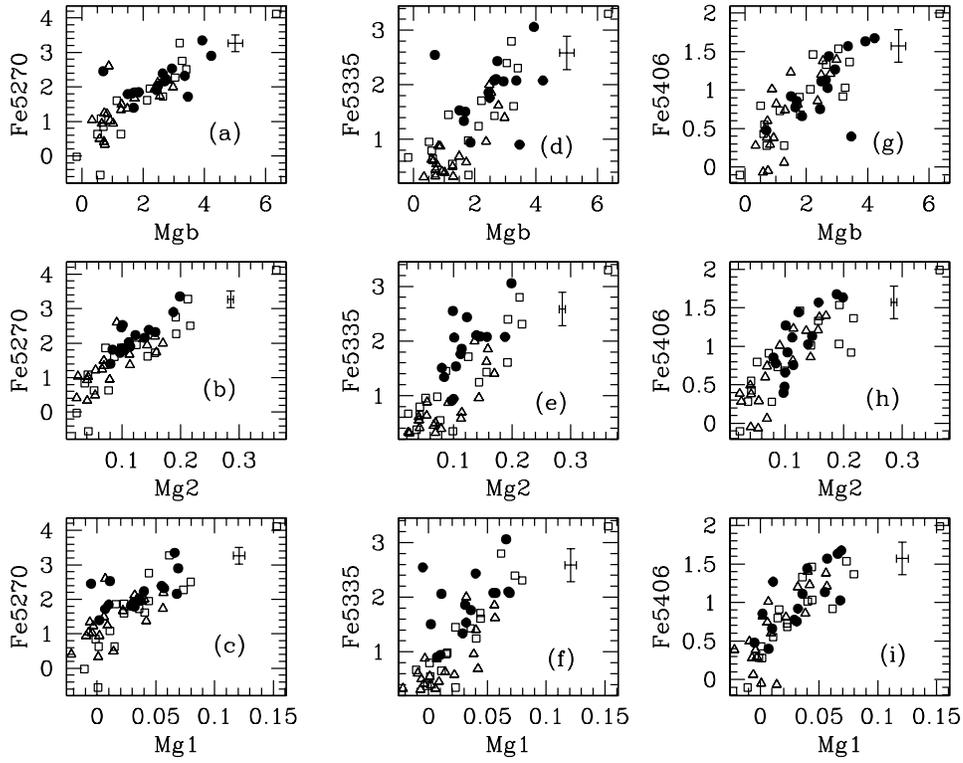}}
\figcaption[iron.eps]{\label{ironfig}Iron indices plotted against magnesium indices
for M81 (filled circles), M31 (open squares) and \MW~(open triangles)
\gcs.  The error bars show the typical sizes of the statistical errors for
the M81 data.  The Fe5335 index appears enhanced, but both Fe5335
and Fe5270 can be affected by the strength of absorption by other
elements \citep{trager98}.  Fe5406 is a pure measure of the iron
abundance, and based on this index we conclude that relative abundances
of magnesium and iron in M81 \gcs~at the metal-rich end of the
metallicity range is consistent with the abundances for \gcs~in M31 and the \MW.
The apparent depression of Mg2 in panel {\it e} of \myfig{elementfig} is
likely due to contamination artificially enhancing the Fe5335 index.}
\end{figure}

\eject
\begin{figure}
\plotone{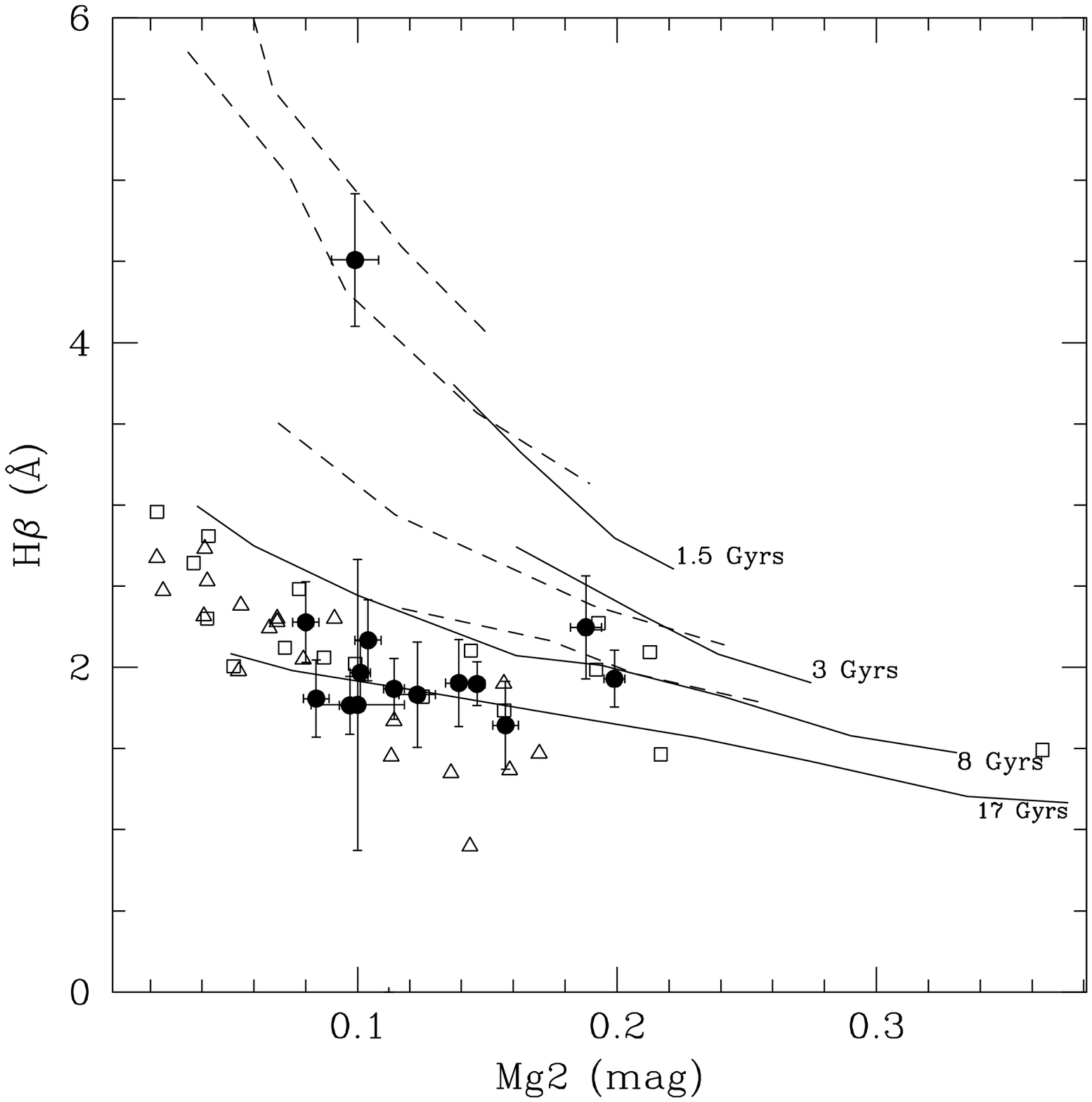}
\figcaption[ages.eps]{\label{agefig} The age-sensitive index \Hbeta~plotted
against the metallicity-sensitive index Mg2 for M81 (filled circles),
M31 (open squares) and \MW~(open triangles) \gcs.  Overplotted as solid lines
are \citet{worthey94} isochrones of ages 1.5 and 3 Gyrs (spanning [Fe/H]
-0.225 to 0.5) and 8 and 17 Gyrs (spanning [Fe/H] of -2.0 to 0.5).
The dashed lines are isochrones of ages 1.5, 3, 8 and 15 Gyrs
from the models of \citet{FB95}, spanning metal-content of \met{-1.3}
to solar.   With the exception of object 15, the M81 cluster ages are
consistent with the ages of M31 and \MW~\gcs, i.e., they are old.}
\end{figure}

\eject
\begin{figure}
\plotone{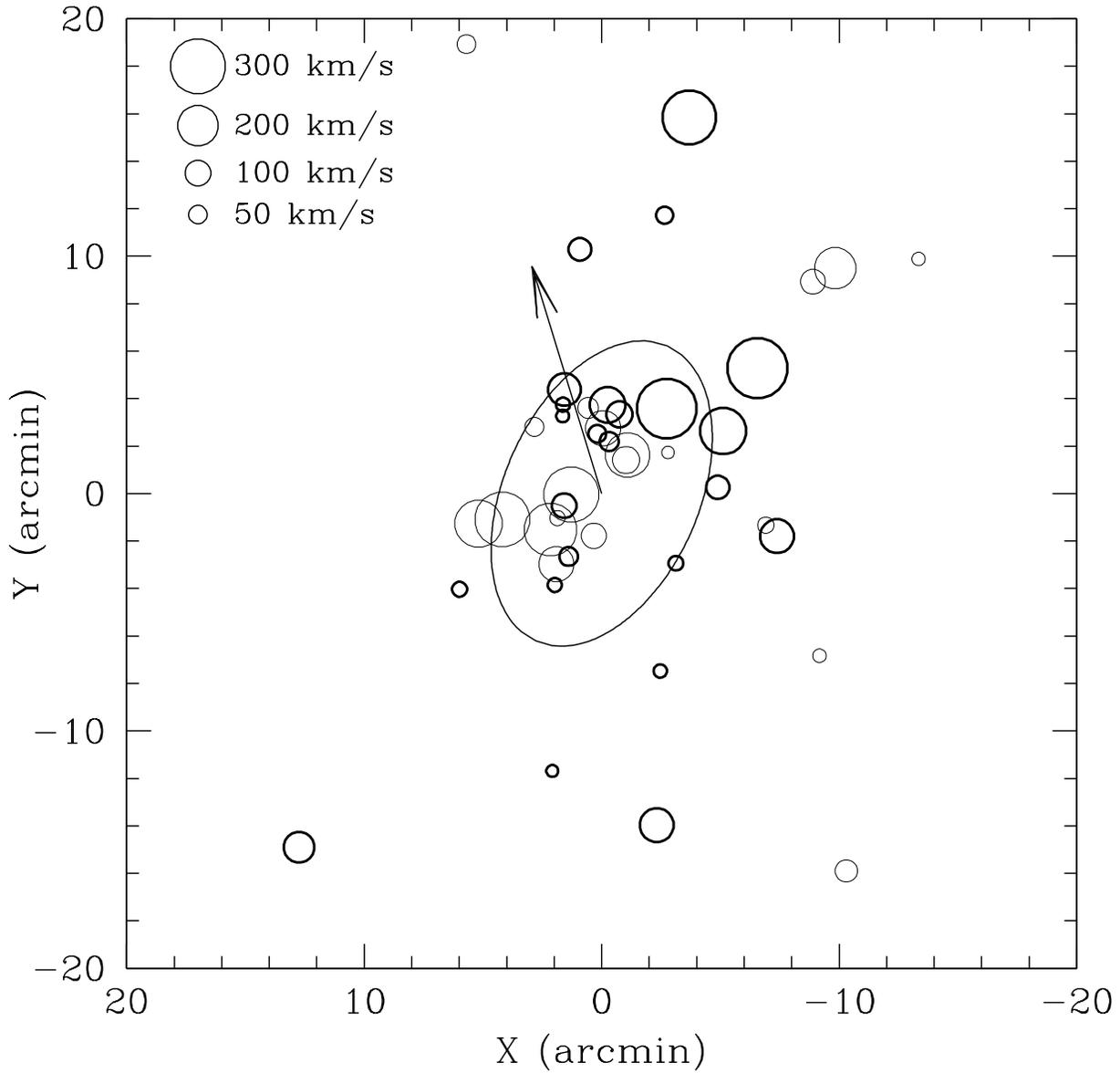}
\figcaption[rotation1.eps]{\label{rot1fig} Velocities and positions of the full
sample of M81 \gcs.  
Thick-lined circles are receding, thin-lined circles are approaching.
The ellipse shows the location of the M81's disk, taken from \citet{PBH95}.  The qualitative impression
is that the M81 cluster system is rotating.  The arrow shows the
rotation axis for the full sample (see \rotsec).}
\end{figure}

\eject
\begin{figure}
\plotone{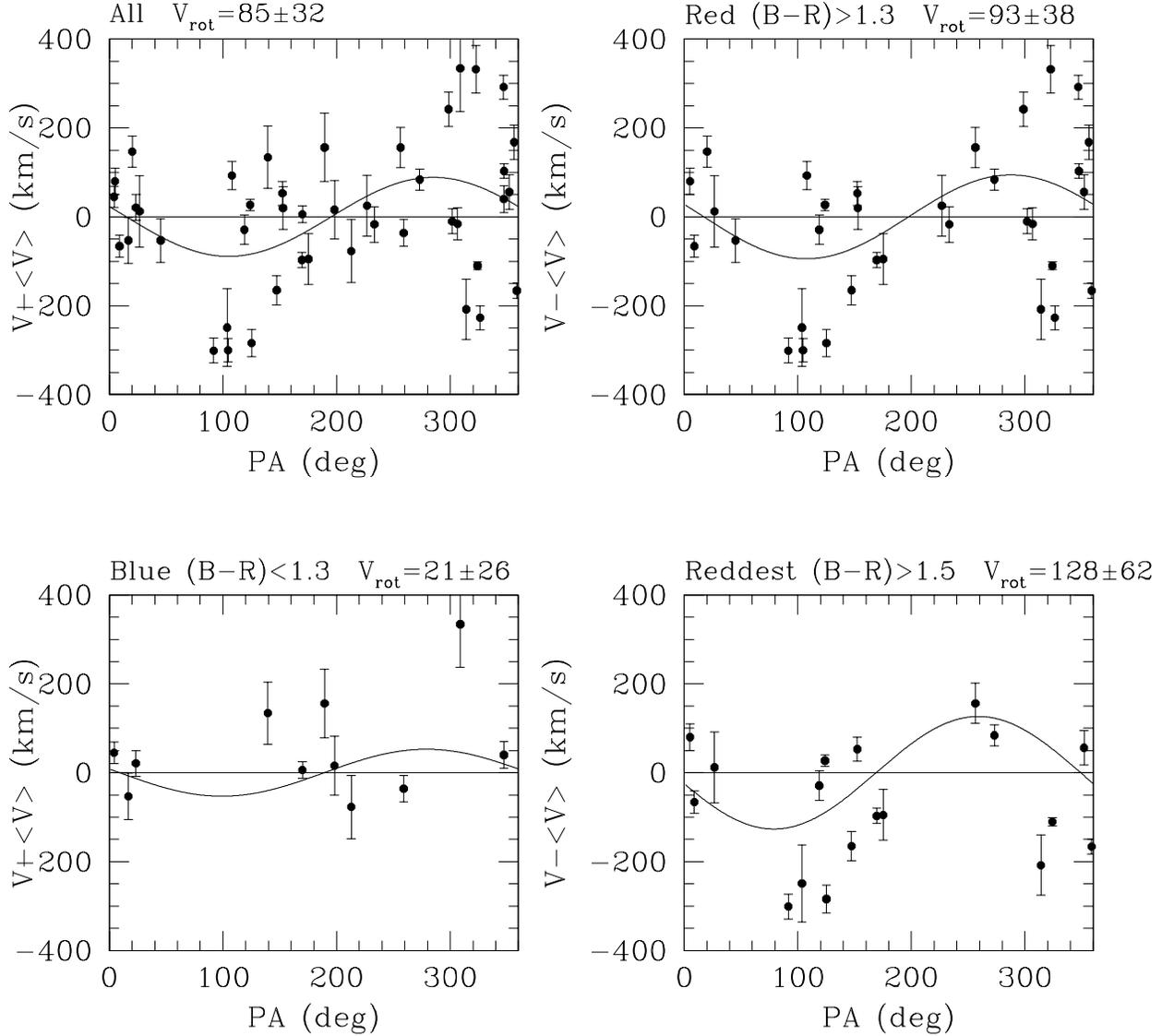}
\figcaption[rot.color.eps]{\label{rotcolorfig} Quantitative assessment of
rotation in M81 \gc~sub-samples based on color.  Panel {\it a}
shows rotation for the full range of colors. 
The rotation velocities were derived using the method of
\citet{pryor_meylan93}.  The vertical axis is the cluster radial velocity
relative to the mean velocity of the full sample.  The horizontal axis
is the position angle between a cluster and the galaxy center, where a
PA of 0 degrees is directly north-south and positive PA is east of
north.  The best fitting sinusoids to these data
represent solid-body rotation and place a lower limit on the
rotation velocity of each sample.  Rotation is strongly suggested
in both of the red, metal-rich sub-samples.  The reddest sub-sample
also has the lowest velocity dispersion (see \mytab{kintable}). The
blue, metal-poor sub-sample shows no significant evidence for
rotation.}
\end{figure}

\begin{figure}
\plotone{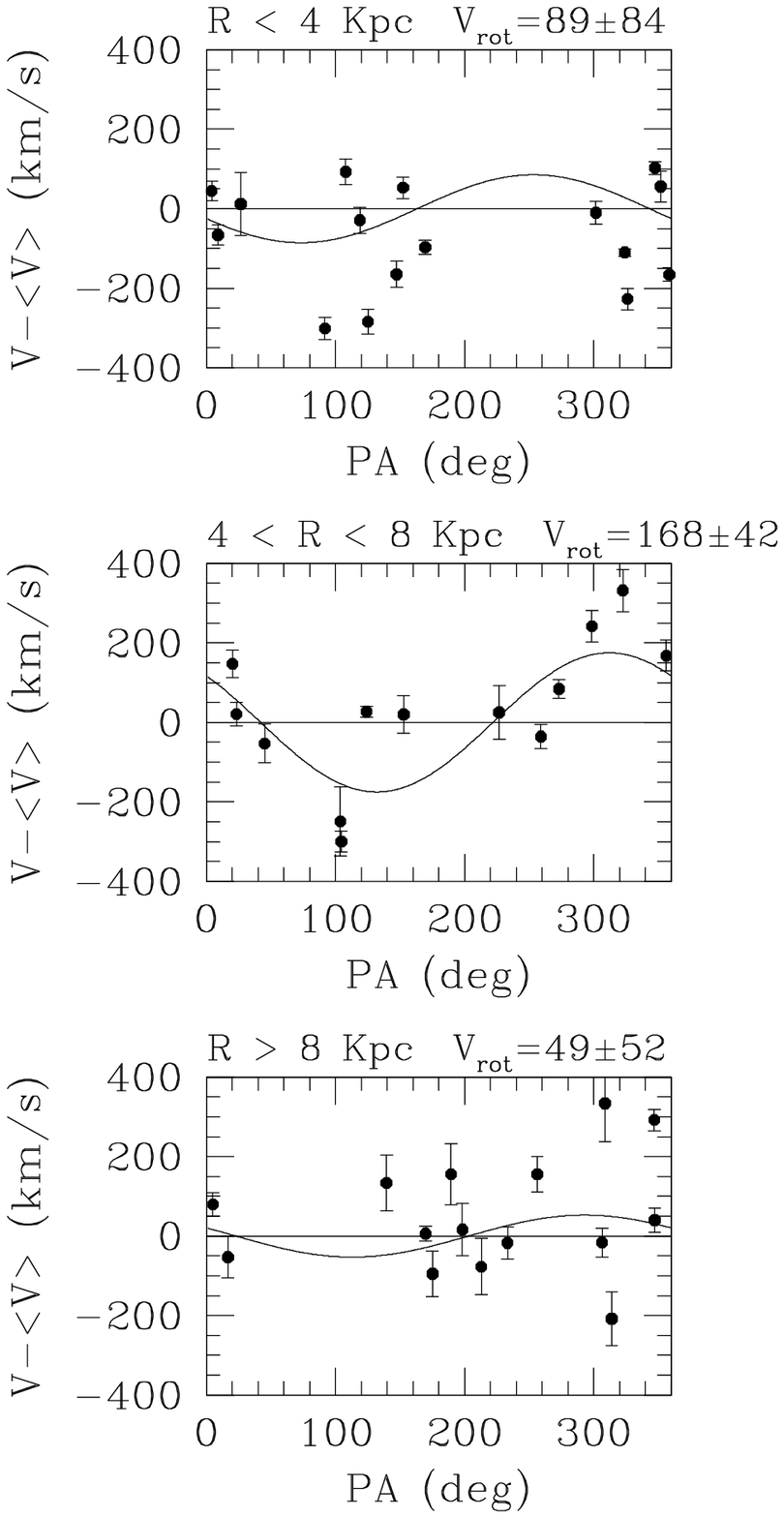}
\figcaption[rot.space.eps]{\label{rotspacefig} Quantitative assessment
of rotation in M81 \gc~sub-samples based on \pgcr.  Only the
sub-sample of clusters with intermediate \pgcri~show evidence for rotation.
See \rotsec.}
\end{figure}

\eject
\begin{figure}
\plotone{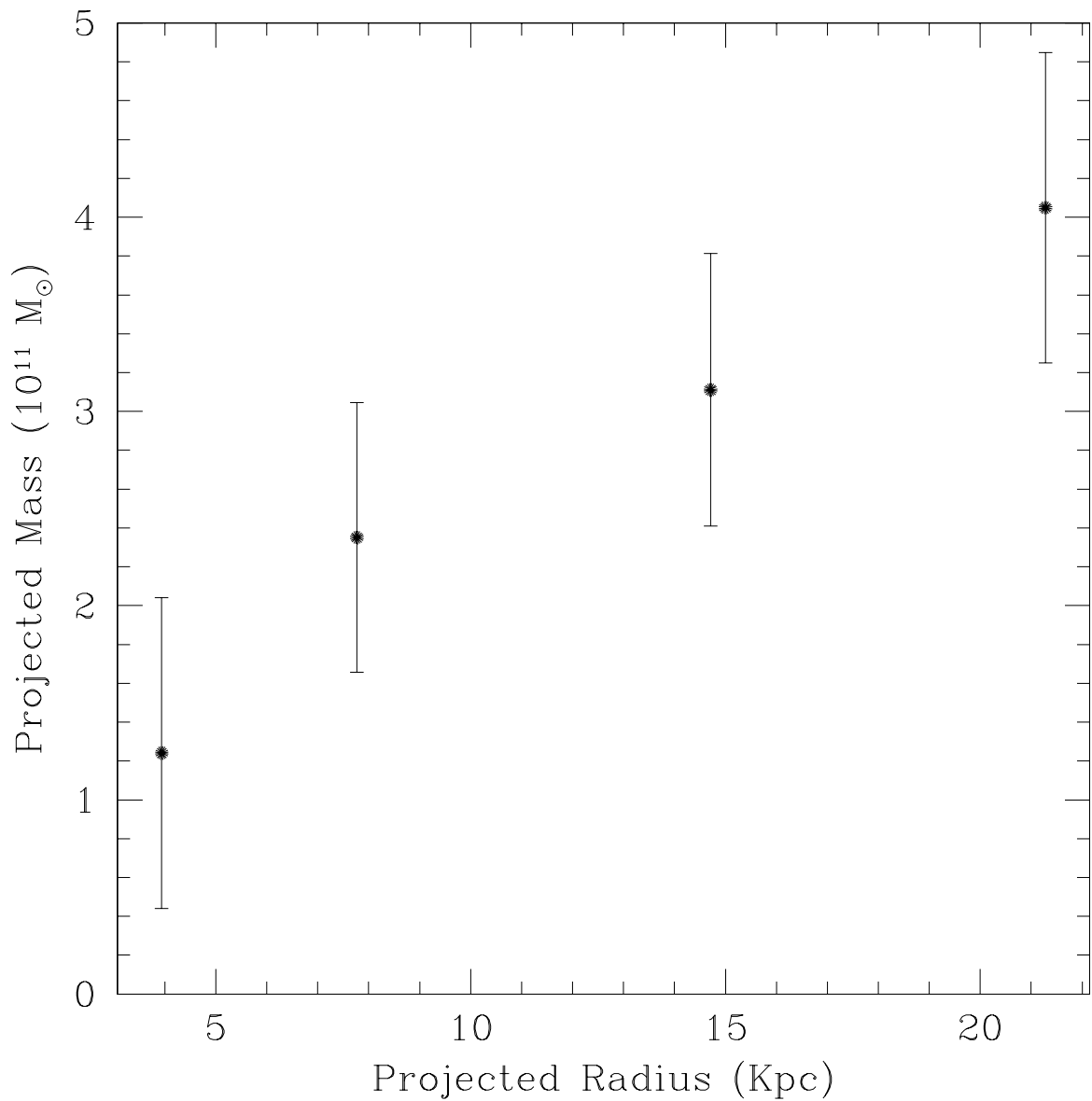}
\figcaption[mass.eps]{\label{massfig} Projected mass vs.~projected
galactocentric radius for M81, derived using radial velocities of M81
\gcs~(see \masssec).  The increase in mass at large radius is
consistent with the presence of a dark-matter halo.}
\end{figure}
\end{onecolumn}
\end{document}